\documentclass[10pt]{article}
\usepackage[utf8]{inputenc}
\usepackage{geometry}
\usepackage{mathtools}
\usepackage{amsfonts}
\usepackage{amsthm}
\newtheorem{proposition}{Proposition}[section]
\newtheorem{corollary}{Corollary}[section]
\allowdisplaybreaks

\makeatletter
\g@addto@macro\normalsize{
  \setlength\abovedisplayskip{\baselineskip}
  \setlength\belowdisplayskip{\baselineskip}
  \setlength\abovedisplayshortskip{\baselineskip}
  \setlength\belowdisplayshortskip{\baselineskip}
  \setlength{\jot}{\baselineskip}
}
\makeatother
\newcommand{\A}{\vphantom{\kappa_{y}^{2}\kappa_{z}^{2}\overline{c}_{\nu}^{2}\underline{c}_{\nu}^{2}}}
\usepackage[dvipsnames]{xcolor}
\usepackage{graphicx}
\usepackage{tikz}
\usepackage{pgfplots}
\usepackage[nolists]{endfloat}

\usepackage[UKenglish]{datetime}
\newdateformat{Currentdate}{%
\THEDAY\ \monthname[\THEMONTH], \THEYEAR}
\usepackage{titlesec}
\titleformat*{\section}{\normalfont\normalsize\bfseries}
\titleformat*{\subsection}{\normalfont\normalsize\bfseries}
\titleformat*{\subsubsection}{\normalfont\normalsize\bfseries}
\titleformat*{\paragraph}{\normalfont\normalsize\bfseries}
\titleformat*{\subparagraph}{\normalfont\normalsize\bfseries}
\usepackage[sort]{natbib}
\setcitestyle{citesep={,}}
\usepackage[capitalise,noabbrev]{cleveref}
\crefname{proposition}{Proposition}{Propositions}
\crefname{corollary}{Corollary}{Corollaries}
\crefname{proof}{Proof}{Proofs}
\title{The $f$-divergence of a von Mises-Fisher distribution from some reference distributions\thanks{%
The authors extend their thanks to Yu-Lin Lin, Hugo Lopez and Debbie Yang for helpful discussion and suggestions. 
The authors also gratefully acknowledge financial support from the ESRC, via the ESRC Centre for Microdata Methods and Practice (CeMMAP; grant number RES-589-28-0001) and an ESRC studentship, respectively. 
Kitagawa further acknowledges financial support from the European Research Council (Starting grant No. 715940).}}
\author{{Toru Kitagawa}\thanks{Brown University, Department of Economics. Email: toru\_kitagawa@brown.edu.} 
\and {Jeff Rowley}\thanks{University College London, Department of Economics.}}
\begin{document}
\maketitle
\begin{abstract}
\noindent
The von Mises-Fisher family is a parametric family of distributions on the surface of the unit ball, summarised by a concentration parameter and a mean direction.
As a quasi-Bayesian prior, a von Mises-Fisher distribution is a convenient and parsimonious choice when parameter spaces are isomorphic to the hypersphere (e.g., maximum score estimation in semi-parametric discrete choice, estimation of single-index treatment assignment rules via empirical welfare maximisation, under-identifying linear simultaneous equation models).
Despite a long history of application, measures of statistical divergence have not been analytically characterised for von Mises-Fisher distributions.
This paper provides analytical expressions for the $f$-divergence of a von Mises-Fisher distribution from another, distinct, von Mises-Fisher distribution in $\mathbb{R}^p$ and from the uniform distribution on the hypersphere.
This paper also collects several other results pertaining to the von Mises-Fisher family of distributions, and characterises the limiting behaviour of the measures of divergence that we consider. 
\vspace{\baselineskip}
\end{abstract}
\section{The von Mises-Fisher family of distributions}
\label{fDIV:SEC-VMF}
The von Mises-Fisher family of distributions is well-known in the field of directional statistics\footnote{%
See \citet{MardiaJupp} for a summary of important results in the field of directional statistics.}
but is foreign to economics and, as such, warrants some introduction.
Also known as the Langevin family \citep{Watson-Langevin}, the von Mises-Fisher family recognises those two titans of statistics, Sir Ronald Fisher and Richard von Mises, for their seminal contributions in considering Gaussianity on the circle \citep{vonMises} and on the sphere \citep{Fisher}. 
Subsequent work has generalised the von Mises-Fisher family to $\mathbb{R}^{p}$, and has led to the definition of other related parametric distributions such as the Bingham family \citep{Bingham} and the Fisher-Bingham or Kent family \citep{Kent}.

A von Mises-Fisher distribution assigns probability mass to the surface of the unit ball -- the hypersphere.
As such, the von Mises-Fisher family is relevant to situations where the researcher is interested in either the sampling of directional vectors -- i.e., vectors of unit length -- or in the clustering of some phenomenon on a circular object, such as occurs if data is periodic.
Applications range from the study of sea turtle navigation \citep{Hillen}, to the study of perihelia of long-tailed comets \citep{MardiaRoyalSociety} and near-earth objects \citep{Sei}, as well as to the study of patient arrival data \citep{MardiaRoyalSociety}.
\citet{Sabelfeld} even links the von Mises-Fisher family to the solving of high-dimensional diffusion-advection-reaction equations.
The von Mises-Fisher family is a two-parameter family, summarised by a concentration parameter (or, simply, concentration), which we denote by $\kappa>0$, and a mean direction, which we denote by $\boldsymbol{\mu}\in\mathbb{R}^{p}$ and which is of unit length.

The main contribution of this paper is to provide analytical expressions for the $f$-divergence of a von Mises-Fisher distribution from two relevant reference distributions given several common choices of function.
We study the broad class of Rényi divergence of \textit{simple order} as well as several other measures of (statistical) divergence that relate to the Rényi class -- the $\chi$-square distance, the squared-Hellinger distance and the Kullback-Leibler divergence.
Each is, of course, a measure of the difference between two probability distributions.
Several well-known inequalities relate these measures to the total variation distance \citep{BretagnolleHuber,Pinsker}, which is often of interest.
The reference distributions that we specify are another, distinct, von Mises-Fisher distribution in $\mathbb{R}^{p}$ and the uniform distribution on the hypersphere.
We are unaware of such expressions being available elsewhere.
Alongside these expressions, we characterise how the various measures of divergence that we consider change as a von Mises-Fisher distribution becomes increasingly concentrated and degenerate.
In particular, we provide asymptotic expansions that complement results given in \citet{KitagawaLopez}, which utilise Hankel expansions and which also rely on results in \citet{Amos}. 
These asymptotic expansions offer analytically tractable polynomial approximations of each of the measures of divergence that we consider in terms of the concentration parameter, with these approximations accurate when this parameter takes large values.

Obtaining analytical expressions of statistical divergence is useful for implementing minimum distance-type or penalised estimation methods, and also for characterising the statistical performance of these procedures. 
See, for instance, \citet{KitagawaLopez}, which builds upon the analytical expression of the Kullback-Leibler divergence that is derived in this paper to estimate the randomised treatment assignment rule that minimises a penalised empirical welfare criterion. 

\begin{figure}
\begingroup\centering
\caption{von Mises-Fisher distributions on the circle}
\vspace{\the\baselineskip}
\label{Circle}
\includegraphics{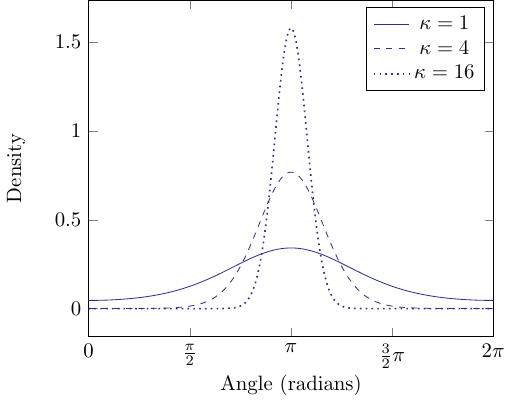}
\\
\endgroup
The density function of a von Mises-Fisher distribution on the circle with mean direction $(-1,0)$ -- i.e., a polar orientation with reference angle equal to $\pi$ radians -- for several values of the concentration parameter. 
A spherical coordinate system is used.
\\
\end{figure}

\begin{figure}
\begingroup\centering
\caption{von Mises-Fisher distributions on the sphere}
\vspace{\the\baselineskip}
\label{Sphere}
\includegraphics{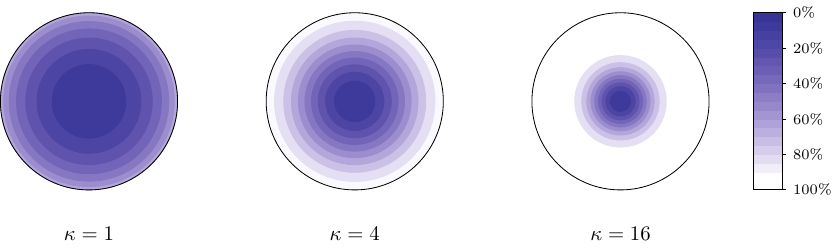}
\\
\endgroup
Orthogonal projection of the sphere, oriented to the mean direction, for several values of the concentration parameter.
Each contour describes a region in which the von Mises-Fisher distribution assigns 10\% mass, with contours distinguished by their shading. 
As the value of the concentration parameter is increased, more mass is assigned to the vicinity of the pole.
\\
\end{figure}

An intention of this paper is to provide an extensive list of analytical expressions for the moments and other known distributional features of the von Mises-Fisher family.
With this objective in mind, we collect several results pertaining to the von Mises-Fisher family that are, to varying extents, available elsewhere.
These include statement of the first two moments of a von Mises-Fisher distribution (available in or adaptable from \citealp[\S 9.3 and \S 9.6, respectively]{MardiaJupp}) and its associated Fisher information matrix \citep{HornikGrun}.
\citet{Hillen,HornikGrun,DhillonSra} demonstrate three distinct approaches to obtaining expressions for these moments.
These are integration by substitution following transformation to spherical coordinates, application of the divergence theorem,\footnote{%
The divergence theorem relates the area of a surface integral to a volume integral.} 
and differentiation of the moment-generating function, respectively.
We use several of the results in these papers directly, whilst others are included for reference only. 
In particular, knowledge of the first moment is essential to characterising the divergence of a von Mises-Fisher distribution from our chosen reference distributions.

Directional objects are common in economics, and the von Mises-Fisher family of distributions is relevant to many environments and several methods.
For instance, consider the canonical binary choice model with latent random utility. 
The rational choice of the individual, which we denote by $D_{i}\in\{0,1\}$, is determined according to the linear index equation 
\begin{equation}
D_{i} 
= 
1(\mathbf{X}_{i}' \boldsymbol{\beta}-U_{i}\geq0),
\end{equation}
where $\mathbf{X}_{i}\in\mathbb{R}^{p}$ and $U_{i}\in\mathbb{R}^{p}$ denote the individual's observable characteristics (including an intercept) and latent heterogeneity, respectively.
The conditional zero-median restriction inherent in the semiparametric maximum score approach of \citet{Manski_1975, Manski_1985}, which translates here as $\mathrm{median}(U_{i}|\mathbf{X}_{i}) = 0$, does not identify the scale of the utility coefficients $\boldsymbol{\beta}\in\mathbb{R}^{p}$.
It is common to normalise the parameter space of $\boldsymbol{\beta}$ to the collection of vectors satisfying $\| \boldsymbol{\beta} \|_{2} = 1$ -- i.e., to the hypersphere defined by the collection of vectors with unit Euclidean length.
Similarly, in the context of statistical treatment choice \citep{manski2004statistical},  \citet{KitagawaTetenov} considers individualised treatment assignment rules based upon a linear index,
\begin{equation}
D_{i} 
= 
1(\mathbf{X}_{i}'\boldsymbol{\beta}\geq 0),
\end{equation}
where $\mathbf{X}_{i}'\boldsymbol{\beta}\in\mathbb{R}$ is an eligibility score that aggregates the individual's observable characteristics and determines whether she should be assigned to treatment -- i.e., if $X_{i}\in\mathbb{R}^{p}$ maps to $D_{i} = 1$ -- or to non-treatment -- i.e., if $X_{i}\in\mathbb{R}^{p}$ maps to $D_{i}=0$. 
Such assignment rules are invariant to multiplication of the eligibility score by a positive constant and can be uniquely indexed by a parameter vector on the hypersphere.

Optimising maximum score or an empirical welfare criterion is difficult, however, and complicates estimation of and inference on $\boldsymbol{\beta}$.
This motivates a quasi-Bayesian approach as considered in \citet{Chernozhukov2003}.
The von Mises-Fisher family offers a parsimonious and convenient prior specification for $\boldsymbol{\beta}$ within the quasi-Bayesian framework, with prior elicitation facilitated by knowledge of the moments of the distribution.
In a related but different context, PAC-Bayesian analysis, which is widely studied in machine learning \citep[to name but a few relevant papers]{McAllester2003,catoni2007pac,Germain2009,Alquier2016}, considers exponentiated negative empirical risk as a quasi-likelihood, and forms a posterior distribution over prediction rules. 
The von Mises-Fisher family is then not only useful as a specified prior over directional parameters, but can also be used to approximate a posterior distribution for $\boldsymbol{\beta}$ in the linear classification rule or linear index treatment assignment rule settings \citep{KitagawaLopez}. 

Another context where the parameter space is isomorphic to the hypersphere is the class of underidentifying linear simultaneous equation models in which the imposed model restrictions identify structural parameters up to sets of orthonormal transformations. 
See, for instance, \citet{Uhlig_2005,Arias_Rubio-Ramirez_Waggoner_2018,Giacomini_Kitagawa_2020a} for a class of set-identified structural vector autoregressions in which the identified set of an impulse-response is spanned by the class of orthonormal matrices. 
The isomorphism of the hypersphere and the orthogonal group suggests that a spherical distribution -- and a von Mises-Fisher distribution in particular -- can be used as a prior distribution for the non-identified orthonormal matrices. 
When combined with prior elicitation of the reduced-form parameters, the moments of a von Mises-Fisher distribution can be used to translate a belief about the structural parameters into a prior distribution over the non-identified orthonormal matrices.
Like the Gaussian family of distributions on the hyperplane from which it can be derived, the von Mises-Fisher family is highly restrictive\footnote{%
The von Mises-Fisher family is akin to the class of Gaussian distributions that feature a diagonal variance matrix -- i.e., statistically independent Gaussian random variables -- with constant entries on the diagonal.} 
but, nonetheless, forms an interesting baseline case to study.

We are unaware of any paper that characterises the $f$-divergence of a von Mises-Fisher distribution as we do.
For instance, \citet{Diethe} similarly studies the Kullback-Leibler divergence of von Mises-Fisher distributions.
Whereas we provide exact analytical expressions, \citet{Diethe} either provides upper bounds on the Kullback-Leibler divergence, or else provides analytical expressions that rely on an approximation that is valid only when the von Mises-Fisher distribution is close to the uniform distribution over the hypersphere, something that we do not rely on.
Where \textit{updating} of the von Mises-Fisher distribution has been considered, this appears to have mainly centred on the likelihood function and its characterisation rather than on measures of divergence \textit{per se}.
We refer to \citet{MardiaEl,Linetal} as pertinent examples. 

A von Mises-Fisher distribution constitutes a conjugate prior \citep{MardiaEl}.
Our choice of reference distributions -- another, distinct, von Mises-Fisher distribution in $\mathbb{R}^{p}$ and the uniform distribution on the hypersphere -- reflects both this and the prevalence of the uniform distribution in practice.
Another commonly invoked choice that we do not consider is the Jeffreys prior, which is proportional to the square root of the determinant of the Fisher information matrix relative to the parametrisation employed.\footnote{%
Our parametrisation of the von Mises-Fisher family specifies a concentration and a mean direction -- what \citet{HornikGrun} call a polar coordinate parametrisation.
Equally, one could combine the concentration and mean direction into a single parameter vector, $\boldsymbol{\eta}$ say, such that the concentration corresponds to $\|\boldsymbol{\eta}\|_{2}$ and the mean direction corresponds to $\boldsymbol{\eta}/\|\boldsymbol{\eta}\|_{2}$.
This is the parametrisation that \citet{HornikGrun} specifies.} 
\citet{HornikGrun} derive the Fisher information matrix and its determinant and show that the Jeffreys prior is improper in this setting.
\section{The probability density function and moments of the von Mises-Fisher family}
\label{fDIV:SEC-VMFMOMENTS}
Throughout this paper, we exploit the fact that the von Mises-Fisher family is an exponential family and, accordingly, we adopt the terminology that is used in conjunction with that well-known class.
Defining $\nu\doteq p/2-1$ for convenience and maintaining $\kappa>0$ and $\|\boldsymbol{\mu}\|_{2}=1$, we write the probability density function of a von Mises-Fisher random vector, which we denote by $\mathbf{x}\in\mathbb{R}^{p}$, for integer $p>1$ and satisfying $\|\mathbf{x}\|_{2}=1$, as 
\begin{equation}
\label{fDIV:EQ-PDF}
f_{p}\left(\mathbf{x};\kappa,\boldsymbol{\mu}\right)
\doteq
\frac{\exp\left(\kappa\boldsymbol{\mu}'\mathbf{x}\right)}{C_{\nu}\left(\kappa\right)},
\end{equation} 
where we reiterate that $\kappa$ and $\boldsymbol{\mu}$ are the concentration and mean direction, respectively, and where
\begin{equation}
\label{fDIV:EQ-CONSTANT}
C_{\nu}\left(\kappa\right)
\doteq
\frac{\left(2\pi\right)^{\nu+1}I_{\nu}\left(\kappa\right)}{\kappa^{\nu}}
=
\int_{\mathbb{S}^{p-1}}\exp\left(\kappa\boldsymbol{\mu}'\mathbf{x}\right)\,\mathrm{d}\mathbf{x},
\end{equation}
thereby guaranteeing that the density function integrates to one.\footnote{%
See \citet[\S B.2]{DhillonSra} for proof of this statement, and also for derivation of the determinant of the Jacobian matrix in the generalised spherical coordinate transform of this integral.} 
We use $I_{\nu}\left(z\right) : \mathbb{R}^2\rightarrow\mathbb{R}$ to denote the modified Bessel function of the first kind (see \cref{fDIV:APPENDIX}, for further details; hereafter, where we say modified Bessel function, we intend this to mean the modified Bessel function of the first kind), and $\mathbb{S}^{p}$ to denote the $p$-sphere -- the collection of unit vectors in $\mathbb{R}^{p+1}$.
We refer to the exponentiation -- i.e., the numerator in \cref{fDIV:EQ-PDF} -- as the kernel of the density function, and to the normalising constant -- i.e., the denominator in \cref{fDIV:EQ-PDF} -- as the partition function.
We emphasise that the integral in \cref{fDIV:EQ-CONSTANT} is over the hypersphere and it is this fact that makes derivation of statistical features of the von Mises-Fisher random vector difficult.
Moreover, we note that our choice of parametrisation is but one way that a von Mises-Fisher distribution can be parametrised.
Another parametrisation that is better suited to certain analyses of von Mises-Fisher distributions (in particular, derivation of their moments) is presented in \citet{HornikGrun}.

The von Mises-Fisher family is the hyperspherical analogue of the Gaussian family, which is informative as to its shape.
This relationship is shown by appropriately normalising the probability density function of statistically independent Gaussian random variables with variance $1/\kappa$ that are distributed on the hypersphere,
\begin{equation}
\left(\int_{\mathbb{S}^{p-1}}\frac{\kappa^{\nu+1}\exp\left(-\kappa\left(\mathbf{x}-\boldsymbol{\mu}\right)'\left(\mathbf{x}-\boldsymbol{\mu}\right)/2\right)}{\left(2\pi\right)^{\nu+1}}\,\mathrm{d}\mathbf{x}\right)^{-1}\frac{\kappa^{\nu+1}\exp\left(-\kappa\left(\mathbf{x}-\boldsymbol{\mu}\right)'\left(\mathbf{x}-\boldsymbol{\mu}\right)/2\right)}{\left(2\pi\right)^{\nu+1}}
=
f_{p}\left(\mathbf{x};\kappa,\boldsymbol{\mu}\right).
\end{equation}
We plot the density function and contours of the cumulative distribution function for several values of the concentration parameter and a common orientation in \cref{Circle,Sphere} for the circular case (when $\mathbb{R}^{p}=\mathbb{R}^{2}$) and the spherical case (when $\mathbb{R}^{p}=\mathbb{R}^{3}$), respectively.
As is to be expected, the von Mises-Fisher family is unimodal and symmetric about its mean direction, with the concentration parameter determining the degeneracy (when $\kappa\rightarrow\infty$) and uniformity (when $\kappa\rightarrow 0$) of the distribution, and it assigns positive density to the entirety of the hypersphere.
Importantly, the von Mises-Fisher family is rotationally invariant, which is the hyperspherical analogue of the location invariance property that is exhibited by the Gaussian family.

The von Mises-Fisher family of distributions coincides with the von Mises and Fisher families in the circular and spherical cases, respectively. 
It is this coincidence that explains the nomenclature.

As explained in \citet{MardiaJupp}, the first moment of the von Mises-Fisher family is in the interior of the unit ball, something which is also true for other named directional distributions -- i.e., it is not on the hypersphere.
Specifically, the first moment of the von Mises-Fisher family has the form
\begin{equation}
\mathbb{E}\left(\mathbf{x}\right)
=
\rho\boldsymbol{\mu},
\end{equation}
where $\rho$ is the (population) mean resultant length and satisfies $0<\rho<1$.
As such, providing a general characterisation of the first and, to a lesser extent, higher-order moments of the von Mises-Fisher family then amounts to characterising the constant of proportionality, the mean resultant length.

We obtain the (centred) moments of the von Mises-Fisher family by differentiating its moment-generating function.
Since the von Mises-Fisher family is an exponential family, the moment-generating function is equal to the log-partition function.
Whilst the partition function of the von Mises-Fisher family has a closed-form expression, this is not necessarily true for other named directional families.
For instance, we are unaware of any closed-form expressions for the partition functions of both the Kent and von Mises families of distributions beyond the bivariate case.\footnote{%
Whilst numerical methods can be used to compute the moments of the Kent and von Mises families (see \citealp{KentGaneiber,Best} for suitable rejection sampling routines; also \citealp{Mardia}, which establishes that the conditional von Mises density is itself von Mises), the lack of closed-form expressions for these other named directional families certainly limits their tractability and usefulness in applications featuring a high-dimensional covariate or outcome vector.}

\begin{proposition}
\label{fDIV:COR-EV}
Given $\mathbf{x}\in\mathbb{S}^{p-1}$ is distributed as a $p$-variate von Mises-Fisher random vector with concentration $\kappa>0$ and mean direction $\boldsymbol{\mu}\in\mathbb{S}^{p-1}$, and recalling that $\nu\doteq p/2-1$, its first and second moments are
\begin{equation}
\mathbb{E}\left(\mathbf{x}\right)
=
r_{\nu}\left(\kappa\right)\boldsymbol{\mu},
\end{equation}
and
\begin{align}
\mathrm{Variance}\left(\mathbf{x}\right)
&=
r_{\nu}\left(\kappa\right)\left(\frac{1}{\kappa}\mathbb{I}_{p}+\left(r_{\nu+1}\left(\kappa\right)-r_{\nu}\left(\kappa\right)\right)\boldsymbol{\mu}\boldsymbol{\mu}'\right),\\
&=
\frac{1}{\kappa}r_{\nu}\left(\kappa\right)\mathbb{I}_{p}+\left(1-\frac{p}{\kappa}r_{\nu}\left(\kappa\right)-r_{\nu}^{2}\left(\kappa\right)\right)\boldsymbol{\mu}\boldsymbol{\mu}',
\end{align}
respectively, where, for comparability with \citet{Amos},
\begin{equation}
r_{\nu}\left(\kappa\right)\doteq \frac{I_{\nu+1}\left(\kappa\right)}{I_{\nu}\left(\kappa\right)},
\end{equation}
which is the ratio of consecutive modified Bessel functions.
\end{proposition}

Although we defer proof to \cref{fDIV:PROOFS}, we emphasise that \cref{fDIV:COR-EV} is adapted from results available elsewhere \citep{MardiaJupp}.
We also note that \citet{MardiaJupp} discuss several other important results aside from the first and second moments, including the asymptotic and high-concentration behaviour of von Mises-Fisher random vectors and their tangent normal vectors.
We emphasise that the expression for the variance that we present here is distinct from the circular variance, which is simply the distance of the mean resultant length from the surface -- i.e., one minus the mean resultant length.
Moreover, and as noted in \citet{HornikGrun}, given that the von Mises-Fisher family is an exponential family, its second moment coincides with its Fisher information matrix under the parametrisation considered in that paper. 

Although we do not present expressions for the higher-order moments of the von Mises-Fisher family, these can be obtained via iterative differentiation.
We present several results in \cref{fDIV:PROOFS} that factor in such a process, and which are central to proving \cref{fDIV:COR-EV}.
These expressions and the modified Bessel functions that underpin them can be implemented in several common statistical programming languages for small to moderately large values of the concentration parameter, with recursive expressions appearing in \citet{Amos} facilitating implementation for larger values. 
\section{Measuring the $f$-divergence of an \textit{obtained distribution} from a \textit{reference distribution}}
\label{fDIV:SEC-VMFDIVERGENCES}
The $f$-divergence measures the divergence of one probability distribution from another.
In what follows, we suppose that 
\begin{gather}
\label{fDIV:EQ-ROTATION}
\begin{aligned}
\mathbf{y}
\sim&
f_{p}(\mathbf{y};\kappa_{y},\boldsymbol{\mu}_{y}),\\
\mathbf{z}
\sim&
f_{p}(\mathbf{z};\kappa_{z},\boldsymbol{\mu}_{z}),
\end{aligned}
\end{gather}
such that $\mathbf{y}\in\mathbb{S}^{p-1}$ and $\mathbf{z}\in\mathbb{S}^{p-1}$ are two von Mises-Fisher random vectors satisfying all of the usual properties that we maintain.\footnote{%
A necessary condition that we require is that the probability distribution of the random vector $\mathbf{y}$ is absolutely continuous with respect to the probability distribution of the random vector $\mathbf{z}$ \citep[\S 2.7 and \S 14.5]{Lattimore}, thereby guaranteeing that each of the measures that we consider is well-defined.
The von Mises-Fisher family clearly satisfies this requirement, assigning positive density to all points on the hypersphere.}
We use the notation above to distinguish the two von Mises-Fisher random vectors from the $\kappa$-concentrated $\boldsymbol{\mu}$-oriented random vector $\mathbf{x}$ that is discussed in the previous sections.

We adopt the convention of referring to the probability distribution of the random vector $\mathbf{y}$ as the \textit{obtained distribution} and to the probability distribution of the random vector $\mathbf{z}$ as the \textit{reference distribution}, and of measuring the divergence of the \textit{obtained distribution} from the \textit{reference distribution}. 

We reiterate that the uniform distribution on the hypersphere corresponds to the limiting case where the concentration of the von Mises-Fisher distribution is zero.
As such, we emphasise that our framework is also compatible with the reference distribution being equal to the uniform distribution on the hypersphere.

For all of the measures that we consider, each measure approaches its maximum possible value when the obtained distribution is degenerate.
We use Bachmann-Landau notation (otherwise known as Big O notation) to characterise the limiting behaviour of these measures with respect to the concentration.
We defer proof of each proposition in this section to \cref{fDIV:PROOFS}.
\subsection{Rényi divergence}
Following \citet{vanErven}, we define the Rényi divergence of \textit{simple order} $\alpha\in\left(0,1\right)\cup\left(1,\infty\right)$ as
\begin{equation}
d_{\alpha}\left(\mathbf{y},\mathbf{z}\right)
\doteq
\frac{1}{\alpha-1}\ln\left(\int_{\mathbb{S}^{p-1}}f_{p}\left(\mathbf{x};\kappa_{y},\boldsymbol{\mu}_{y}\right)^{\alpha} f_{p}\left(\mathbf{x};\kappa_{z},\boldsymbol{\mu}_{z}\right)^{1-\alpha}\,\mathrm{d}\mathbf{x}\right).
\end{equation}
The Rényi divergence is the most general measure of $f$-divergence that we consider and describes a broad class that relates to the other measures that we study -- namely, the $\chi$-squared distance, squared-Hellinger distance and Kullback-Leibler divergence.

\begin{proposition}
\label{fDIV:COR-RD}
Given $\mathbf{y}\in\mathbb{S}^{p-1}$ and $\mathbf{z}\in\mathbb{S}^{p-1}$ are distributed as $p$-variate von Mises-Fisher random vectors with concentrations $\kappa_{y}>0$ and $\kappa_{z}>0$, respectively, and mean directions $\boldsymbol{\mu}_{y}\in\mathbb{S}^{p-1}$ and $\boldsymbol{\mu}_{z}\in\mathbb{S}^{p-1}$, respectively, and recalling that $\nu\doteq p/2-1$, the Rényi divergence of simple order $\alpha\in\left(0,1\right)\cup\left(1,\infty\right)$ is
\begin{equation}
\label{fDIV:EQ-RENYI1}
d_{\alpha}\left(\mathbf{y},\mathbf{z}\right)
=
\frac{\nu}{\alpha-1}\ln\left(\frac{\kappa_{y}^{\alpha}\kappa_{z}^{1-\alpha}}{\kappa_{\alpha}}\right)+\frac{\alpha}{\alpha-1}\ln\left(\frac{I_{\nu}\left(\kappa_{\alpha}\right)}{I_{\nu}\left(\kappa_{y}\right)}\right)-\ln\left(\frac{I_{\nu}\left(\kappa_{\alpha}\right)}{I_{\nu}\left(\kappa_{z}\right)}\right),
\end{equation}
where
\begin{equation}
\label{fDIV:EQ-RENYIWEIGHTING}
\kappa_{\alpha}
\doteq
\left\|\alpha\kappa_{y}\boldsymbol{\mu}_{y}+\left(1-\alpha\right)\kappa_{z}\boldsymbol{\mu}_{z}\right\|_{2}.
\end{equation}
In the special case where $\kappa_{\alpha}=0$, 
\begin{equation}
\label{fDIV:EQ-RENYI2}
d_{\alpha}\left(\mathbf{y},\mathbf{z}\right)
=
\frac{\nu}{\alpha-1}\ln\left(\frac{\kappa_{y}^{\alpha}\kappa_{z}^{1-\alpha}}{2}\right)+\frac{\alpha}{\alpha-1}\ln\left(\frac{1/\Gamma\left(\nu+1\right)}{I_{\nu}\left(\kappa_{y}\right)}\right)-\ln\left(\frac{1/\Gamma\left(\nu+1\right)}{I_{\nu}\left(\kappa_{z}\right)}\right),
\end{equation}
with
\begin{equation}
\label{fDIV:EQ-RENYISPECCONDITIONS}
\kappa_{z}
=
\frac{\alpha}{\left\vert 1-\alpha\right\vert}\kappa_{y}
\text{ and }
\boldsymbol{\mu}_{z}+\mathrm{Sign}\left(1-\alpha\right)\boldsymbol{\mu}_{y}=0.
\end{equation}
In the special case where $\mathbf{z}$ is, instead, uniformly distributed on the hypersphere,
\begin{equation}
\label{fDIV:EQ-RENYI3}
d_{\alpha}\left(\mathbf{y},\mathbf{z}\right)
=
\frac{\nu}{\alpha-1}\ln\left(\frac{2^{1-\alpha}}{\alpha\kappa_{y}^{1-\alpha}}\right)+\frac{\alpha}{\alpha-1}\ln\left(\frac{I_{\nu}\left(\alpha\kappa_{y}\right)}{I_{\nu}\left(\kappa_{y}\right)}\right)-\ln\left(\frac{I_{\nu}\left(\alpha\kappa_{y}\right)}{1/\Gamma\left(\nu+1\right)}\right).
\end{equation}
The Rényi divergence is an $\mathrm{O}\left(\ln\left(\kappa_{y}\right)\right)$ function.
\end{proposition}
\subsection{$\chi$-square distance}
We define the $\chi$-square distance as
\begin{align}
d_{\chi}\left(\mathbf{y},\mathbf{z}\right)
&\doteq
\int_{\mathbb{S}^{p-1}}\frac{\left(f_{p}\left(\mathbf{y};\kappa_{y},\boldsymbol{\mu}_{y}\right)-f_{p}\left(\mathbf{z};\kappa_{z},\boldsymbol{\mu}_{z}\right)\right)^{2}}{f_{p}\left(\mathbf{z};\kappa_{z},\boldsymbol{\mu}_{z}\right)}\,\mathrm{d}\mathbf{x},\\
&=\label{fDIV:EQ-CSD}
\int_{\mathbb{S}^{p-1}}\frac{f_{p}\left(\mathbf{y};\kappa_{y},\boldsymbol{\mu}_{y}\right)^{2}}{f_{p}\left(\mathbf{z};\kappa_{z},\boldsymbol{\mu}_{z}\right)}\,\mathrm{d}\mathbf{x}-1,
\end{align}
with \cref{fDIV:EQ-CSD} being the more convenient definition to work with.
We observe that the $\chi$-square distance relates to the Rényi class of measures via the equality relation 
\begin{equation}
\label{fDIV:EQ-RENYITOCSD}
d_\alpha\left(\mathbf{y},\mathbf{z}\right)|_{\alpha=2}=\ln\left(1+d_{\chi}\left(\mathbf{y},\mathbf{z}\right)\right),
\end{equation}
as per \citet{vanErven}, which serves to illustrate the breadth of the Rényi class.

\begin{proposition}
\label{fDIV:COR-CSD}
Given $\mathbf{y}\in\mathbb{S}^{p-1}$ and $\mathbf{z}\in\mathbb{S}^{p-1}$ are distributed as $p$-variate von Mises-Fisher random vectors with concentrations $\kappa_{y}>0$ and $\kappa_{z}>0$, respectively, and mean directions $\boldsymbol{\mu}_{y}\in\mathbb{S}^{p-1}$ and $\boldsymbol{\mu}_{z}\in\mathbb{S}^{p-1}$, respectively, and recalling that $\nu\doteq p/2-1$, the $\chi$-square distance is
\begin{equation}
\label{fDIV:EQ-CHISQ1}
d_{\chi}\left(\mathbf{y},\mathbf{z}\right)
=
\frac{\kappa_{y}^{2\nu}I_{\nu}\left(\kappa_{\chi}\right)I_{\nu}\left(\kappa_{z}\right)}{\kappa_{\chi}^{\nu}\kappa_{z}^{\nu}I_{\nu}\left(\kappa_{y}\right)^{2}}-1,
\end{equation}
where
\begin{equation}
\kappa_{\chi}
\doteq
\left\|2\kappa_{y}\boldsymbol{\mu}_{y}-\kappa_{z}\boldsymbol{\mu}_{z}\right\|_{2}.
\end{equation}
In the special case where $\kappa_{\chi}=0$, 
\begin{equation}
\label{fDIV:EQ-CHI}
d_{\chi}\left(\mathbf{y},\mathbf{z}\right)
=
\frac{\kappa_{y}^{\nu}I_{\nu}\left(2\kappa_{y}\right)}{4^{\nu}\Gamma\left(\nu+1\right)I_{\nu}\left(\kappa_{y}\right)^{2}}-1,
\end{equation}
with
\begin{equation}
\kappa_{z}
=
2\kappa_{y}
\text{ and }
\boldsymbol{\mu}_{z}-\boldsymbol{\mu}_{y}=0.
\end{equation}
In the special case where $\mathbf{z}$ is, instead, uniformly distributed on the hypersphere, the $\chi$-square distance coincides with \cref{fDIV:EQ-CHI}.
The $\chi$-square distance is an $\mathrm{O}\left(\kappa_{y}\right)$ function.
\end{proposition}
\subsection{Squared-Hellinger distance}
We define the squared-Hellinger distance as
\begin{align}
d_{h}\left(\mathbf{y},\mathbf{z}\right)^{2}
\doteq&
\int_{\mathbb{S}^{p-1}}\left(\sqrt{f_{p}\left(\mathbf{x};\kappa_{y},\boldsymbol{\mu}_{y}\right)}-\sqrt{f_{p}\left(\mathbf{x};\kappa_{z},\boldsymbol{\mu}_{z}\right)}\right)^{2}\,\mathrm{d}\mathbf{x},\\
={}&\label{fDIV:EQ-HD}
2\left(1-\int_{\mathbb{S}^{p-1}}\sqrt{f_{p}\left(\mathbf{x};\kappa_{y},\boldsymbol{\mu}_{y}\right)f_{p}\left(\mathbf{x};\kappa_{z},\boldsymbol{\mu}_{z}\right)}\,\mathrm{d}\mathbf{x}\right),
\end{align}
with \cref{fDIV:EQ-HD} being the more convenient definition to work with.
We observe that the squared-Hellinger distance relates to the Rényi class of measures via the equality relation 
\begin{equation}
\label{fDIV:EQ-RENYITOHEL}
d_\alpha\left(\mathbf{y},\mathbf{z}\right)|_{\alpha=\frac{1}{2}}=-2\ln\left(1-\frac{d_{h}\left(\mathbf{y},\mathbf{z}\right)^{2}}{2}\right),
\end{equation}
as per \citet{vanErven}, which is also interpretable as twice the negative logarithm of the Bhattacharrya coefficient.
The Bhattacharyya coefficient is an approximate measure of the amount of overlap between two probability distributions, such that when the obtained and reference distributions are close (i.e., they have a similar concentration and mean direction) then the squared-Hellinger distance is small in absolute value, as is to be expected.

\begin{proposition}
\label{fDIV:COR-HD}
Given $\mathbf{y}\in\mathbb{S}^{p-1}$ and $\mathbf{z}\in\mathbb{S}^{p-1}$ are distributed as $p$-variate von Mises-Fisher random vectors with concentrations $\kappa_{y}>0$ and $\kappa_{z}>0$, respectively, and mean directions $\boldsymbol{\mu}_{y}\in\mathbb{S}^{p-1}$ and $\boldsymbol{\mu}_{z}\in\mathbb{S}^{p-1}$, respectively, and recalling that $\nu\doteq p/2-1$, the squared-Hellinger distance is
\begin{equation}
\label{fDIV:EQ-HEL1}
d_{h}\left(\mathbf{y},\mathbf{z}\right)^{2}
=
2\left(1-\sqrt{\frac{\kappa_{y}^{\nu}\kappa_{z}^{\nu}I_{\nu}^{2}\left(\kappa_{h}\right)}{\kappa_{h}^{2\nu}I_{\nu}\left(\kappa_{y}\right)I_{\nu}\left(\kappa_{z}\right)}}\right),
\end{equation}
where we define
\begin{equation}
\kappa_{h}
\doteq
\frac{1}{2}\left\|\kappa_{y}\boldsymbol{\mu}_{y}+\kappa_{z}\boldsymbol{\mu}_{z}\right\|_{2}.
\end{equation}
In the special case where $\kappa_{h}=0$,
\begin{equation}
\label{fDIV:EQ-HEL2}
d_{h}\left(\mathbf{y},\mathbf{z}\right)^{2}
=
2\left(1-\frac{\kappa_{y}^{\nu}}{2^{\nu}I_{\nu}\left(\kappa_{y}\right)\Gamma\left(\nu+1\right)}\right),
\end{equation}
with
\begin{equation}
\kappa_{z}
=
\kappa_{y}
\text{ and }
\boldsymbol{\mu}_{z}+\boldsymbol{\mu}_{y}=0.
\end{equation}
In the special case where $\mathbf{z}$ is, instead, uniformly distributed on the hypersphere,
\begin{equation}
\label{fDIV:EQ-HEL3}
d_{h}\left(\mathbf{y},\mathbf{z}\right)^{2}
=
2\left(1-\sqrt{\frac{2^{3\nu}I_{\nu}^{2}\left(\kappa_{y}/2\right)\Gamma\left(\nu+1\right)}{\kappa_{y}^{\nu}I_{\nu}\left(\kappa_{y}\right)}}\right).
\end{equation}
The squared-Hellinger distance is an $\mathrm{O}\left(1-1/\kappa_{y}\right)$ function.
\end{proposition}
\subsection{Kullback-Leibler divergence}
We define the Kullback-Leibler divergence as 
\begin{equation}
\label{fDIV:EQ-KL}
d_{\ell}\left(\mathbf{y},\mathbf{z}\right)
\doteq
\int_{\mathbb{S}^{p}}\ln\left(\frac{f_{p}\left(\mathbf{x};\kappa_{y},\boldsymbol{\mu}_{y}\right)}{f_{p}\left(\mathbf{x};\kappa_{z},\boldsymbol{\mu}_{z}\right)}\right)f_{p}\left(\mathbf{x};\kappa_{y},\boldsymbol{\mu}_{y}\right)\,\mathrm{d}\mathbf{x}.
\end{equation}
We observe that the Kullback-Leibler divergence is a limiting case of the Rényi divergence. 
That is,
\begin{equation}
\label{fDIV:EQ-ORDERLIMIT}
\lim_{\alpha\rightarrow1}d_\alpha\left(\mathbf{y},\mathbf{z}\right)=d_\ell\left(\mathbf{y},\mathbf{z}\right),
\end{equation}
as per \citet{vanErven}.

\begin{proposition}
\label{fDIV:COR-KL}
Given $\mathbf{y}\in\mathbb{S}^{p-1}$ and $\mathbf{z}\in\mathbb{S}^{p-1}$ are distributed as $p$-variate von Mises-Fisher random vectors with concentrations $\kappa_{y}>0$ and $\kappa_{z}>0$, respectively, and mean directions $\boldsymbol{\mu}_{y}\in\mathbb{S}^{p-1}$ and $\boldsymbol{\mu}_{z}\in\mathbb{S}^{p-1}$, respectively, and recalling that $\nu\doteq p/2-1$, the Kullback-Leibler divergence is
\begin{equation}
\label{fDIV:EQ-KL1}
d_{\ell}\left(\mathbf{y},\mathbf{z}\right)
=
\nu\ln\left(\frac{\kappa_{y}}{\kappa_{z}}\right)-\ln\left(\frac{I_{\nu}\left(\kappa_{y}\right)}{I_{\nu}\left(\kappa_{z}\right)}\right)+r_{\nu}\left(\kappa_{y}\right)\left(\kappa_{y}\boldsymbol{\mu}_{y}-\kappa_{z}\boldsymbol{\mu}_{z}\right)'\boldsymbol{\mu}_{y}.
\end{equation}
In the special case where $\mathbf{z}$ is, instead, uniformly distributed on the hypersphere,
\begin{equation}
\label{fDIV:EQ-KL2}
d_{\ell}\left(\mathbf{y},\mathbf{z}\right)
=
\nu\ln\left(\frac{\kappa_{y}}{2}\right)-\ln\left(I_{\nu}\left(\kappa_{y}\right)\right)-\ln\left(\Gamma\left(\nu+1\right)\right)+r_{\nu}\left(\kappa_{y}\right)\kappa_{y}.
\end{equation}
The Kullback-Leibler divergence is an $\mathrm{O}\left(\ln\left(\kappa_{y}\right)\right)$ function.
\end{proposition}

We note that the final terms of \cref{fDIV:EQ-KL1,fDIV:EQ-KL2} are proportional to the first moment of a von Mises-Fisher distribution (specifically, the first moment of the obtained distribution).
In particular, we note that the final term of \cref{fDIV:EQ-KL1} can otherwise be written as 
\begin{equation}
\label{fDIV:EQ-EDIFF1}
r_{\nu}\left(\kappa_{y}\right)\left(\kappa_{y}\boldsymbol{\mu}_{y}-\kappa_{z}\boldsymbol{\mu}_{z}\right)'\boldsymbol{\mu}_{y}
=
r_{\nu}\left(\kappa_{y}\right)\left(\kappa_{y}-\kappa_{z}\boldsymbol{\mu}_{z}'\boldsymbol{\mu}_{y}\right),
\end{equation}
which we contrast with the corresponding term in \cref{fDIV:EQ-KL2}.
The difference between \cref{fDIV:EQ-EDIFF1,fDIV:EQ-KL2} arises because the mean direction does not enter the probability density function of the uniform distribution on the hypersphere. 
The intuition here is that, in the special case where the reference distribution is the uniform distribution on the hypersphere, the mean direction can always be taken to be equal to the mean direction of the obtained distribution such that the information gain is interpretable as learning about the concentration only.
In the more general case where the reference distribution is another, distinct, von Mises-Fisher distribution, the information gain is interpretable as learning about the concentration and the mean direction depending upon their respective values. 
\subsection{Total variation distance}
We define the total variation distance, for all measurable $\mathbf{A}\subset \mathbb{S}^{p-1}$, as
\begin{equation}
d_{t}\left(\mathbf{y},\mathbf{z}\right)
\doteq
\sup_{\mathbf{A}}\left\vert\int_{\mathbf{A}} f_{p}\left(\mathbf{x};\kappa_{y},\boldsymbol{\mu}_{y}\right)-f_{p}\left(\mathbf{x};\kappa_{z},\boldsymbol{\mu}_{z}\right)\,\mathrm{d}\mathbf{x}\right\vert,
\end{equation}
which is well-defined if the underlying probability space is endowed with the Borel $\sigma$-algebra -- something that we, implicitly, maintain throughout all parts of our analysis.

\begin{corollary}
\label{fDIV:COR-TV}
Given $\mathbf{y}\in\mathbb{S}^{p-1}$ and $\mathbf{z}\in\mathbb{S}^{p-1}$ are distributed as $p$-variate von Mises-Fisher random vectors with concentrations $\kappa_{y}>0$ and $\kappa_{z}>0$, respectively, and mean directions $\boldsymbol{\mu}_{y}\in\mathbb{S}^{p-1}$ and $\boldsymbol{\mu}_{z}\in\mathbb{S}^{p-1}$, respectively, and recalling that $\nu\doteq p/2-1$, the total variation distance satisfies the inequality relations
\begin{align}
d_{t}\left(\mathbf{y},\mathbf{z}\right)^{2}\leq d_{h}\left(\mathbf{y},\mathbf{z}\right)^{2}\leq d_{\ell}\left(\mathbf{y},\mathbf{z}\right)\leq d_{\chi}\left(\mathbf{y},\mathbf{z}\right),\\
d_{t}\left(\mathbf{y},\mathbf{z}\right)\leq\sqrt{1-\exp\left(-d_{\ell}\left(\mathbf{y},\mathbf{z}\right)\right)}\leq1-\frac{1}{2}d_{\ell}\left(\mathbf{y},\mathbf{z}\right),
\end{align}
and, for all orders $\alpha\in\left(0,1\right]$,
\begin{equation}
\label{fDIV:EQ-PINSKER}
d_t\left(\mathbf{y},\mathbf{z}\right)^2\leq\frac{\alpha}{2}d_\alpha\left(\mathbf{y},\mathbf{z}\right),
\end{equation}
where each measure of divergence is as defined in \cref{fDIV:SEC-VMFDIVERGENCES}.
\end{corollary}

The inequalities in \cref{fDIV:COR-TV} are well-known (see, for instance, \citealp[\S 14.3 and references therein]{Lattimore}, and \citealp{vanErven}).
The total variation distance is difficult to characterise in this setting due to the need to integrate over a region of the hypersphere, which is non-trivial.
The inequality relations stated in \cref{fDIV:COR-TV} provide a means to bound the total variation distance from above using measures that are more easily characterised.
In particular, \cref{fDIV:EQ-PINSKER} is a generalisation of Pinsker's inequality, yielding the classic statement of that inequality at the limit (see \cref{fDIV:EQ-ORDERLIMIT}).
\section{Limiting behaviour of the modified Bessel function and its ratios}
\label{fDIV:SEC-LIMITING}
We now discuss the limiting behaviour of the modified Bessel function and its ratios as a von Mises-Fisher distribution becomes increasingly concentrated around its mean direction.
Such discussion is useful, in particular, for understanding how the various measures of divergence that we consider respond to changes -- specifically, an increase -- in the value of the concentration parameter.

We begin by referring to \citet{Amos} and introducing several results that are stated therein, adapting the notation of that paper to suit our purposes.
Firstly, \citet{Amos} shows that, for all $\kappa_{y}\geq\kappa_{z}>0$,
\begin{equation}
\label{fDIV:EQ-AMOSI}
L_{\nu}\left(\kappa_{y},\kappa_{z}\right)\leq\ln\left(I_{\nu}\left(\kappa_{y}\right)\right)\leq U_{\nu}\left(\kappa_{y},\kappa_{z}\right),
\end{equation}
where
\begin{equation}
\label{fDIV:EQ-L}
L_{\nu}\left(\kappa_{y},\kappa_{z}\right)
\doteq
\ln\left(I_{\nu}\left(\kappa_{z}\right)\right)+\nu\ln\left(\frac{\kappa_{y}}{\kappa_{z}}\right)
+
\underline{c}_{\nu}\ln\left(\frac{\underline{c}_{\nu}+\sqrt{\A\kappa_{z}^{2}+\overline{c}_{\nu}^{2}}}{\underline{c}_{\nu}+\sqrt{\A\kappa_{y}^{2}+\overline{c}_{\nu}^{2}}}\right)
+
\frac{\kappa_{y}^{2}-\kappa_{z}^{2}}{\sqrt{\A\kappa_{y}^{2}+\overline{c}_{\nu}^{2}}+\sqrt{\A\kappa_{z}^{2}+\overline{c}_{\nu}^{2}}},
\end{equation}
and
\begin{equation}
\label{fDIV:EQ-U}
U_{\nu}\left(\kappa_{y},\kappa_{z}\right)
\doteq
\ln\left(I_{\nu}\left(\kappa_{z}\right)\right)
+
\nu\ln\left(\frac{\kappa_{y}}{\kappa_{z}}\right)+\underline{c}_{\nu}\ln\left(\frac{\underline{c}_{\nu}+\sqrt{\A\kappa_{z}^{2}+\underline{c}_{\nu}^{2}}}{\underline{c}_{\nu}+\sqrt{\A\kappa_{y}^{2}+\underline{c}_{\nu}^{2}}}\right)
+
\frac{\kappa_{y}^{2}-\kappa_{z}^{2}}{\sqrt{\A\kappa_{y}^{2}+\underline{c}_{\nu}^{2}}+\sqrt{\A\kappa_{z}^{2}+\underline{c}_{\nu}^{2}}},
\end{equation}
given $\underline{c}_{\nu}\doteq \nu+1/2$ and $\overline{c}_{\nu}\doteq \nu+3/2$. 
We note that \cref{fDIV:EQ-AMOSI} is reversed when the concentrations of the obtained and reference distributions instead satisfies $\kappa_{z}>\kappa_{y}$.
In the special case where $\kappa_{z}\rightarrow 0$, which we reiterate corresponds to the case where the reference distribution is the uniform distribution on the hypersphere, \citet{Amos} shows that \cref{fDIV:EQ-L,fDIV:EQ-U} reduce to
\begin{equation}
\label{fDIV:EQ-LUNIFORM}
L_{\nu}\left(\kappa_{y},0\right)
=
\frac{1}{2}\ln\left(\frac{2}{\kappa_{y}}\right)
-
\ln\left(\Gamma\left(\nu+1\right)\right)
+
\underline{c}_{\nu}\ln\left(\frac{\kappa_{y}\left(\underline{c}_{\nu}+\overline{c}_{\nu}\right)/2}{\underline{c}_{\nu}+\sqrt{\A\kappa_{y}^{2}+\overline{c}_{\nu}^{2}}}\right)
+
\frac{\kappa_{y}^{2}}{\overline{c}_{\nu}+\sqrt{\A\kappa_{y}^{2}+\overline{c}_{\nu}^{2}}},
\end{equation}
and 
\begin{equation}
\label{fDIV:EQ-UUNIFORM}
U_{\nu}\left(\kappa_{y},0\right)
=
\frac{1}{2}\ln\left(\frac{2}{\kappa_{y}}\right)
-
\ln\left(\Gamma\left(\nu+1\right)\right)
+
\underline{c}_{\nu}\ln\left(\frac{\kappa_{y}\left(\underline{c}_{\nu}+\underline{c}_{\nu}\right)/2}{\underline{c}_{\nu}+\sqrt{\A\kappa_{y}^{2}+\underline{c}_{\nu}^{2}}}\right)
+
\frac{\kappa_{y}^{2}}{\underline{c}_{\nu}+\sqrt{\A\kappa_{y}^{2}+\underline{c}_{\nu}^{2}}},
\end{equation}
respectively.
Secondly, \citet{Amos} shows that
\begin{equation}
\label{fDIV:EQ-AMOSII}
\frac{\kappa_{y}}{\underline{c}_{\nu}+\sqrt{\A\kappa_{y}^{2}+\overline{c}_{\nu}^{2}}}
\leq
r_{\nu}\left(\kappa_{y}\right)
\leq
\frac{\kappa_{y}}{\underline{c}_{\nu}+\sqrt{\A\kappa_{y}^{2}+\underline{c}_{\nu}^{2}}},
\end{equation}
where we again rely on the preceding definitions. 
This inequality defines a subset of the unit interval.

\begin{corollary}
\label{fDIV:COR-LF}
Given that $\kappa>0$, then for $L_{\nu}\left(\kappa,0\right)$ and $U_{\nu}\left(\kappa,0\right)$ defined as in \cref{fDIV:EQ-LUNIFORM,fDIV:EQ-UUNIFORM} and recalling that $\nu\doteq p/2-1$,
\begin{equation}
L_{\nu}\left(\kappa,0\right)
\geq
\kappa-\frac{1}{2}\ln\left(\kappa\right)-\nu\ln\left(2\right)-\ln\left(\Gamma\left(\nu+1\right)\right)-\overline{c}_{\nu},
\end{equation}
and
\begin{equation}
U_{\nu}\left(\kappa,0\right)
\leq
\kappa-\frac{1}{2}\ln\left(\kappa\right)+\frac{1}{2}\ln\left(2\right)-\ln\left(\Gamma\left(\nu+1\right)\right)+\underline{c}_{\nu}\ln\left(\underline{c}_{\nu}\right)
\end{equation}
such that $\ln\left(I_{\nu}\left(\kappa\right)\right)$ is an $\mathrm{O}\left(\kappa-\ln\left(\kappa\right)/2\right)$ function.
\end{corollary}

Although we defer proof to \cref{fDIV:PROOFS}, we emphasise that \cref{fDIV:COR-LF} follows from \cref{fDIV:EQ-LUNIFORM,fDIV:EQ-UUNIFORM} and the simple exploitation of the properties of increasing concave functions.
Moreover, that we choose to state that the logarithm of the modified Bessel function is an $\mathrm{O}\left(\kappa-\ln\left(\kappa\right)/2\right)$ function rather than a linear function is for practical reasons.
Specifically, we rely on \cref{fDIV:COR-LF} to prove many of the statements in \cref{fDIV:SEC-VMFDIVERGENCES}, for which we find that the linear component typically cancels. 
We observe that \cref{fDIV:COR-LF} aligns with results elsewhere \cite[\S 10.30]{DLMF}.

One further quantity that is often of interest is the circular variance of the von Mises-Fisher family of distributions,\footnote{%
See \citealp{MardiaJupp} for further, general, details about the circular variance.} 
which is defined as one minus the ratio of modified Bessel functions -- i.e., one minus the mean resultant length.
\citet{KitagawaLopez} demonstrates that the circular variance is an $\mathrm{O}\left(1/\kappa\right)$ function.
We observe that the same result can also be attained via Hankel series expansion (see \cref{fDIV:EQ-HANKEL} for a definition and \cref{fDIV:HANKEL} for a demonstration).
We note that Hankel series expansion is appropriate when $\kappa\rightarrow\infty$, which is the limiting behaviour that we are interested in.
We emphasise that the circular variance is strictly contained in the unit interval, which is asymptotically guaranteed by the fact that $p\geq 2$ -- i.e., the minimal hypersphere (the circle) is defined on the real plane. 
\section{Conclusions}
The main contribution of this paper is to provide analytical expressions for the $f$-divergence of a von Mises-Fisher distribution from two relevant reference distributions given several common choices of function.
These expressions can be input into several common statistical programming languages to form part of an estimation routine -- for instance, to optimise maximum score or an empirical welfare criterion. 
In characterising the limiting behaviour of our chosen measures of divergence, we provide results that are useful for the theoretical development of econometric methods, especially those that involve penalisation.

We suggest that there are several directions for further research.
Firstly, the von Mises-Fisher family is highly restrictive. 
It would be useful to provide similar characterisations for related parametric distributions such as the Bingham family and the Fisher-Bingham or Kent family that are less restrictive and allow for asymmetry and correlation.
Secondly, directional objects are common in economics but use of the von Mises-Fisher family and related distributions is extremely rare.
We point to several examples where the von Mises-Fisher family would augment or be an attractive choice for existing economic theory, with one example being as a choice of prior in structural vector autoregression.
\numberwithin{equation}{section}
\crefalias{section}{appendix}
\setcounter{section}{0}
\renewcommand*\thesection{\Alph{section}}
\section{Useful functions}
\label{fDIV:APPENDIX}
The modified Bessel function of the first kind is defined in \citet[\S 10.25 and \S 10.32]{DLMF} and, its precursor, \citet[\S 9.6]{AS}, and we refer the reader to those references for further (detailed) information about this function.
The modified Bessel function is defined, for all $z>0$, as
\begin{equation}
\label{fDIV:EQ-BESSELGAMMA}
I_{\nu}\left(z\right)
\doteq
\left(z/2\right)^{\nu}{\displaystyle\sum_{k=0}^\infty}\frac{\left(z/2\right)^{2k}}{k!\Gamma\left(\nu+k+1\right)}
=
\frac{\left(z/2\right)^{\nu}}{\sqrt{\pi}\Gamma\left(\nu+1/2\right)}\int_{0}^{\pi}\exp\left(\pm z\cos\left(\theta\right)\right)\sin^{2\nu}\left(\theta\right)\,\mathrm{d}\theta,
\end{equation}  
or as 
\begin{equation}
\label{fDIV:EQ-BESSELINT}
I_{\nu}\left(z\right)
\doteq
\frac{1}{\pi}\left(\int_{0}^{\pi}\exp\left(z\cos\left(\theta\right)\right)\cos\left(\nu\theta\right)\,\mathrm{d}\theta-\sin\left(\nu\pi\right)\int_{0}^{\infty}\exp\left(-\nu t-z\cosh t\right)\,\mathrm{d}t\right),
\end{equation}
where $\nu$ is said to be the order and $z$ is said to be the argument.
Various other definitions of the modified Bessel function exist, including as the $g : \mathbb{C}\rightarrow\mathbb{R}$ that solves
\begin{equation}
z^{2}\frac{\mathrm{d}^{2}g}{\mathrm{d}z^{2}}+z\frac{\mathrm{d}g}{\mathrm{d}z}-\left(z^{2}+\nu^{2}\right)g=0,
\end{equation}
which is a modification of Bessel's equation. 
We emphasise that the solutions to the above equation are \textit{imaginary}.\footnote{%
Bessel's equation is generally written with the final term on the left-hand side entering additively.
It enters negatively here because we admit purely imaginary solutions.
This is what distinguishes a Bessel function (that admits real solutions) from a modified Bessel function (that admits imaginary solutions).}
Bessel's equation and the modified Bessel function relate to Laplace's equation and harmonic functions (that describe the propagation of a wave along a taut string).
The modified Bessel function satisfies the recurrence relations \cite[\S 10.29]{DLMF}.
\begin{gather}
\label{fDIV:EQ-RECUR}
\begin{aligned}
I_{\nu}\left(z\right)
&=
\frac{z}{2v}\left(I_{\nu-1}\left(z\right)-I_{\nu+1}\left(z\right)\right),\\
I_{\nu}'\left(z\right)
&=
I_{\nu-1}\left(z\right)-\frac{v}{z}I_{\nu}\left(z\right)
=
I_{\nu+1}\left(z\right)+\frac{v}{z}I_{\nu}\left(z\right),
\end{aligned}
\end{gather}
and the limiting behaviour \cite[\S 10.30]{DLMF}
\begin{equation}
\label{fDIV:EQ-LIMITINGFORM} 
\lim_{z\rightarrow 0}\frac{I_{\nu}\left(z\right)}{z^{\nu}}=\frac{1}{2^{\nu}\Gamma\left(v+1\right)}.
\end{equation}
Moreover, the modified Bessel function admits, for all $n\in\mathbb{N}$ as $z\rightarrow\infty$, the Poincaré asymptotic expansion \cite[\S 2.1]{DLMF}
\begin{equation}
\label{fDIV:EQ-HANKEL}
I_{\nu}\left(z\right)
=
\frac{\exp\left(z\right)}{\sqrt{2\pi z}}\left(\sum_{j=0}^{n-1}\left(-1\right)^{j}\frac{a_{j}\left(\nu\right)}{z^{j}}+\mathrm{O}\left(\frac{1}{z^{n}}\right)\right),
\end{equation}
where $\left(a_{j}\left(\nu\right)\right)_{j=0}^{\infty}$ is a sequence of constants satisfying, for all $j\in\mathbb{N}$,
\begin{equation}
\label{fDIV:EQ-HANKELSERIES}
a_{j}\left(\nu\right)\doteq\frac{\prod_{i=1}^{j}\left(4\nu^{2}-\left(2j-i\right)^{2}\right)}{8^{j}j!},
\end{equation}
with $a_{0}\left(\nu\right)=1$.
This sequence of constants is derived from Hankel's expansion \cite[\S 10.40]{DLMF}.
The modified Bessel function takes a particularly useful form in certain instances. 
\section{Proofs}
\label{fDIV:PROOFS}
\begin{proof}[$\hookrightarrow$ Proof of \cref{fDIV:COR-EV}]
We reiterate that the von Mises-Fisher family is an exponential family and, as such, its moments can be obtained via differentiation of the log-partition function, with the first moment equal to the first derivative, the second moment equal to the second derivative, and so forth.
To facilitate our analysis of the log-partition function and its derivatives, we rewrite the probability density function of the von Mises-Fisher family in terms of a single parameter vector.
We let 
\begin{gather}
\begin{aligned}
\kappa
&\doteq
\|\boldsymbol{\eta}\|_2,\\
\boldsymbol{\mu}
&\doteq
\|\boldsymbol{\eta}\|_2^{-1}\boldsymbol{\eta},\\
\end{aligned}
\end{gather}
where $\boldsymbol{\eta}\in\mathbb{R}^{p}$. 
The log-partition function is equal to
\begin{equation}
\label{fDIV:EQ-LOGPARTITION}
\ln\int_{\mathbb{S}^{p-1}}\exp(\kappa\boldsymbol{\mu}'\mathbf{x})\,\mathrm{d}\mathbf{x}
=
\ln(C_{\nu}(\kappa))
=
-\nu\ln\left(\kappa\right)+\left(\nu+1\right)\ln\left(2\pi\right)+\ln\left(I_{\nu}\left(\kappa\right)\right).
\end{equation}
The moments of the von Mises-Fisher family of distributions are obtained by recursively differentiating this function with respect to the new parameter vector. 

We now present some derivatives that are useful for the construction of the first, second and higher-order moments.
Firstly,
\begin{gather}
\label{fDIV:EQ-USEFULDEF1}
\begin{aligned}
\frac{\mathrm{d}}{\mathrm{d}\boldsymbol{\eta}}\kappa
&=
\boldsymbol{\mu},\\
\frac{\mathrm{d}}{\mathrm{d}\boldsymbol{\eta}'}\boldsymbol{\mu}
&=
\frac{1}{\kappa}\left(\mathbb{I}_{p}-\boldsymbol{\mu}\boldsymbol{\mu}'\right).
\end{aligned}
\end{gather}
Secondly, for all $n\in\mathbb{N}_{+}$,
\begin{gather}
\label{fDIV:EQ-USEFULDEF2}
\begin{aligned}
\frac{\mathrm{d}}{\mathrm{d}\boldsymbol{\eta}'}\frac{I_{\nu+n}\left(\kappa\right)}{I_{\nu}\left(\kappa\right)}
&=
\frac{I_{\nu}\left(\kappa\right)I_{\nu+n}'\left(\kappa\right)-I_{\nu+n}\left(\kappa\right)I_{\nu}'(\kappa)}{I_{\nu}\left(\kappa\right)^{2}}\boldsymbol{\mu}',\\
&=
\left(\frac{I_{\nu+n+1}\left(\kappa\right)}{I_{\nu}\left(\kappa\right)}+\frac{nI_{\nu+n}\left(\kappa\right)}{\kappa I_{\nu}\left(\kappa\right)}-\frac{I_{\nu+n}\left(\kappa\right)}{I_{\nu}\left(\kappa\right)}\frac{I_{\nu+1}\left(\kappa\right)}{I_{\nu}\left(\kappa\right)}\right)\boldsymbol{\mu}',\\
&=
\left(\prod_{j=0}^{n-1}r_{\nu+j}\left(\kappa\right)\right)\left(r_{\nu+n}\left(\kappa\right)+\frac{n}{\kappa}-r_{\nu}\left(\kappa\right)\right)\boldsymbol{\mu}',
\end{aligned}
\end{gather}
which exploits \cref{fDIV:EQ-RECUR} and the telescoping property of the ratios.
We now differentiate \cref{fDIV:EQ-LOGPARTITION} (i.e., we take the first derivative of the log-partition function), which yields
\begin{gather}
\label{fDIV:EQ-EPROOF}
\begin{aligned}
\frac{\mathrm{d}}{\mathrm{d}\boldsymbol{\eta}}\ln\left(C_{\nu}\left(\kappa\right)\right)
&=
r_{\nu}\left(\kappa\right)\boldsymbol{\mu},
\end{aligned}
\end{gather}
where we use \cref{fDIV:EQ-USEFULDEF1} and \cref{fDIV:EQ-RECUR}.
This proves the first result of the corollary.
We then differentiate \cref{fDIV:EQ-EPROOF} (i.e., we take the second derivative of the log-partition function), which yields
\begin{equation}
\label{fDIV:EQ-VARIANCE}
\frac{\mathrm{d}}{\mathrm{d}\boldsymbol{\eta}'}r_{\nu}\left(\kappa\right)\boldsymbol{\mu}
=
r_{\nu}\left(\kappa\right)\left(\frac{1}{\kappa}\mathbb{I}_{p}+\left(r_{\nu+1}\left(\kappa\right)-r_{\nu}\left(\kappa\right)\right)\boldsymbol{\mu}\boldsymbol{\mu}'\right),
\end{equation}
where we use \cref{fDIV:EQ-RECUR,fDIV:EQ-USEFULDEF1,fDIV:EQ-USEFULDEF2}.
Substituting 
\begin{equation}
r_{\nu}\left(\kappa\right)r_{\nu+1}\left(\kappa\right)
=
1-\frac{p}{\kappa}r_{\nu}\left(\kappa\right),
\end{equation}
which is valid by \cref{fDIV:EQ-RECUR}, we obtain an alternative expression for the variance.
This proves the second result of the corollary.
The higher-order moments of the von Mises-Fisher family can be obtained via recursive differentiation of these expressions, with application of the product rule of differentiation and the derivatives that are presented in \cref{fDIV:EQ-USEFULDEF1,fDIV:EQ-USEFULDEF2} then sufficient to construct these moments.
\end{proof}
\begin{proof}[$\hookrightarrow$ Proof of \cref{fDIV:COR-RD}]
We focus on the integrand in the definition of the Rényi divergence, and note that
\begin{equation}
\label{fDIV:EQ-RENYIINTEGRAND}
f_{p}\left(\mathbf{x};\kappa_{y},\boldsymbol{\mu}_{y}\right)^{\alpha} f_{p}\left(\mathbf{x};\kappa_{z},\boldsymbol{\mu}_{z}\right)^{1-\alpha}
=
\frac{\exp\left(\left(\alpha\kappa_{y}\boldsymbol{\mu}_{y}+\left(1-\alpha\right)\kappa_{z}\boldsymbol{\mu}_{z}\right)'\mathbf{x}\right)}{C_{\nu}\left(\kappa_{y}\right)^{\alpha} C_{\nu}\left(\kappa_{z}\right)^{1-\alpha}}.
\end{equation}
We then define 
\begin{gather}
\label{fDIV:EQ-KAMUA}
\begin{aligned}
\kappa_{\alpha}
&\doteq
\left\|\alpha\kappa_{y}\boldsymbol{\mu}_{y}+\left(1-\alpha\right)\kappa_{z}\boldsymbol{\mu}_{z}\right\|_{2},\\
\boldsymbol{\mu}_{\alpha}
&\doteq
\left\|\alpha\kappa_{y}\boldsymbol{\mu}_{y}+\left(1-\alpha\right)\kappa_{z}\boldsymbol{\mu}_{z}\right\|_{2}^{-1}\left(\alpha\kappa_{y}\boldsymbol{\mu}_{y}+\left(1-\alpha\right)\kappa_{z}\boldsymbol{\mu}_{z}\right),
\end{aligned}
\end{gather}
such that the resulting expression respects the conventions that we have adopted for the concentration and mean direction -- i.e., that the concentration is non-negative and that the mean direction is a unit vector.
We substitute \cref{fDIV:EQ-KAMUA} into \cref{fDIV:EQ-RENYIINTEGRAND} and integrate over the hypersphere, finding that
\begin{equation}
\label{fDIV:EQ-RENYISIMPLE}
\frac{1}{\alpha-1}\ln\left(\int_{\mathbb{S}^{p-1}}\frac{\exp\left(\kappa_{\alpha}\boldsymbol{\mu}_{\alpha}'\mathbf{x}\right)}{C_{\nu}\left(\kappa_{y}\right)^{\alpha} C_{\nu}\left(\kappa_{z}\right)^{1-\alpha}}\,\mathrm{d}\mathbf{x}\right)
=
\frac{1}{\alpha-1}\ln\left(\frac{C_{\nu}\left(\kappa_{\alpha}\right)^{\alpha} C_{\nu}\left(\kappa_{\alpha}\right)^{1-\alpha}}{C_{\nu}\left(\kappa_{y}\right)^{\alpha} C_{\nu}\left(\kappa_{z}\right)^{1-\alpha}}\right).
\end{equation}
Hence, \cref{fDIV:EQ-RENYIINTEGRAND,fDIV:EQ-RENYISIMPLE} imply that 
\begin{align}
d_{\alpha}\left(\mathbf{y},\mathbf{z}\right)
&=
\frac{\alpha}{\alpha-1}\ln\left(\frac{C_{\nu}\left(\kappa_{\alpha}\right)}{C_{\nu}\left(\kappa_{y}\right)}\right)-\ln\left(\frac{C_{\nu}\left(\kappa_{\alpha}\right)}{C_{\nu}\left(\kappa_{z}\right)}\right),\\
&=
\frac{\alpha}{\alpha-1}\ln\left(\frac{\kappa_{y}^{\nu}I_{\nu}\left(\kappa_{\alpha}\right)}{\kappa_{\alpha}^{\nu}I_{\nu}\left(\kappa_{y}\right)}\right)-\ln\left(\frac{\kappa_{z}^{\nu}I_{\nu}\left(\kappa_{\alpha}\right)}{\kappa_{\alpha}^{\nu}I_{\nu}\left(\kappa_{z}\right)}\right),
\end{align}
which yields \cref{fDIV:EQ-RENYI1} upon rearrangement.

In the special case where $\kappa_{\alpha}=0$, the numerator on the right-hand side of \cref{fDIV:EQ-RENYIINTEGRAND} is one, with the corresponding integral over the hypersphere equal to the surface area of the unit ball.
The formula for the surface area of the unit ball is well-known.
Hence, \cref{fDIV:EQ-RENYISIMPLE} reduces to
\begin{equation}
\frac{1}{\alpha-1}\ln\left(\int_{\mathbb{S}^{p-1}}\frac{1}{C_{\nu}\left(\kappa_{y}\right)^{\alpha} C_{\nu}\left(\kappa_{z}\right)^{1-\alpha}}\,\mathrm{d}\mathbf{x}\right)
=
\frac{1}{\alpha-1}\ln\left(\frac{2\pi^{\nu+1}/\Gamma\left(\nu+1\right)}{C_{\nu}\left(\kappa_{y}\right)^{\alpha} C_{\nu}\left(\kappa_{z}\right)^{1-\alpha}}\right),
\end{equation}
which in turn means that
\begin{equation}
d_{\alpha}\left(\mathbf{y},\mathbf{z}\right)
=
\frac{\nu}{\alpha-1}\ln\left(\frac{\kappa_{y}^{\alpha}\kappa_{z}^{1-\alpha}}{2}\right)
+
\frac{\alpha}{\alpha-1}\ln\left(\frac{1/\Gamma\left(\nu+1\right)}{I_{\nu}\left(\kappa_{y}\right)}\right)
-
\ln\left(\frac{1/\Gamma\left(\nu+1\right)}{I_{\nu}\left(\kappa_{z}\right)}\right).
\end{equation}
We reiterate that this special case occurs only when \cref{fDIV:EQ-RENYISPECCONDITIONS} holds, which facilitates the restatement of the Renyí divergence in terms of one of the two concentration parameters. 
We choose to state the Renyí divergence in terms of both concentration parameters because the expression that we obtain is relatively simple.

In the special case where the reference distribution is the uniform distribution on the hypersphere, the integrand in the definition of the Rényi divergence has the specific form 
\begin{equation}
\label{fDIV:EQ-RENYIUNIFORM}
\lim_{\kappa_{z}\rightarrow 0}f_{p}\left(\mathbf{x};\kappa_{y},\boldsymbol{\mu}_{y}\right)^{\alpha} f_{p}\left(\mathbf{x};\kappa_{z},\boldsymbol{\mu}_{z}\right)^{1-\alpha}
=
\frac{\exp\left(\alpha\kappa_{y}\boldsymbol{\mu}_{y}'\mathbf{x}\right)\Gamma\left(\nu+1\right)^{1-\alpha}}{\left(2\pi^{\nu+1}\right)^{1-\alpha}C_{\nu}\left(\kappa_{y}\right)^{\alpha}},
\end{equation}
which again utilises the formula for the surface area of the unit ball.
As such, \cref{fDIV:EQ-RENYISIMPLE} reduces to
\begin{equation}
\label{fDIV:EQ-RENYIUNIFORMINTEGRAL}
\frac{1}{\alpha-1}\ln\left(\int_{\mathbb{S}^{p-1}}\frac{\exp\left(\alpha\kappa_{y}\boldsymbol{\mu}_{y}'\mathbf{x}\right)\Gamma\left(\nu+1\right)^{1-\alpha}}{\left(2\pi^{\nu+1}\right)^{1-\alpha}C_{\nu}\left(\kappa_{y}\right)^{\alpha}}\,\mathrm{d}\mathbf{x}\right)
=
\frac{1}{\alpha-1}\ln\left(\frac{C_{\nu}\left(\alpha\kappa_{y}\right)\Gamma\left(\nu+1\right)^{1-\alpha}}{\left(2\pi^{\nu+1}\right)^{1-\alpha}C_{\nu}\left(\kappa_{y}\right)^{\alpha}}\right).
\end{equation}
We then note that
\begin{equation}
\label{fDIV:EQ-RENYILOG1}
\frac{1}{\alpha-1}\ln\left(\frac{C_{\nu}\left(\alpha\kappa_{y}\right)^{\alpha}}{C_{\nu}\left(\kappa_{y}\right)^{\alpha}}\right)
=
\frac{\alpha}{\alpha-1}\ln\left(\frac{I_{\nu}\left(\alpha\kappa_{y}\right)}{I_{\nu}\left(\kappa_{y}\right)}\right)-\frac{\alpha}{\alpha-1}\nu\ln\left(\alpha\right),
\end{equation}
and
\begin{equation}
\label{fDIV:EQ-RENYILOG2}
\frac{1}{\alpha-1}\ln\left(\frac{C_{\nu}\left(\alpha\kappa_{y}\right)^{1-\alpha}\Gamma\left(\nu+1\right)^{1-\alpha}}{\left(2\pi^{\nu+1}\right)^{1-\alpha}}\right)
=
-\ln\left(\frac{I_{\nu}\left(\alpha\kappa_{y}\right)}{1/\Gamma\left(\nu+1\right)}\right)-\nu\left(\ln\left(\frac{2}{\kappa_{y}}\right)-\ln\left(\alpha\right)\right).
\end{equation}
\cref{fDIV:EQ-RENYILOG1,fDIV:EQ-RENYILOG2} sum to form \cref{fDIV:EQ-RENYIUNIFORMINTEGRAL}, which we have already established is the formula for the Rényi divergence in this case.
We then obtain \cref{fDIV:EQ-RENYI3} upon rearrangement.

We now turn our attention to the limiting behaviour of the Rényi divergence with respect to an increase in $\kappa_{y}$ given that $\kappa_{y}\geq\kappa_{z}$, holding $\kappa_{z}>0$ fixed.
We rely extensively on \cref{fDIV:COR-LF}, and define 
\begin{equation}
\label{fDIV:EQ-RENYIDIFF}
g_{\alpha}
\doteq
\kappa_{\alpha}-\alpha\kappa_{y}-\left(1-\alpha\right)\kappa_{z},
\end{equation}
for convenience.
Given that 
\begin{equation}
\label{fDIV:EQ-RENYIDIFFEFFECT}
\kappa_{\alpha}
\in
[\alpha\kappa_{y}-|1-\alpha|\kappa_{z},\alpha\kappa_{y}+|1-\alpha|\kappa_{z}],
\end{equation}
so
\begin{equation}
\label{fDIV:EQ-RENYIDIFFIMPLICATION}
g_{\alpha}
\in
[\min\left(0,2\left(\alpha-1\right)\kappa_{z}\right),\max\left(0,2\left(\alpha-1\right)\kappa_{z}\right)],
\end{equation}
thereby establishing that the Rényi-weighted difference in concentrations is bounded.
We recall that
\begin{gather}
\label{fDIV:EQ-BOUNDDEFINITION}
\begin{aligned}
L_{\nu}\left(\kappa,0\right)
&\geq 
\kappa-\frac{1}{2}\ln\left(\kappa\right)-\underline{g}\left(\nu\right),\\
U_{\nu}\left(\kappa,0\right)
&\leq 
\kappa-\frac{1}{2}\ln\left(\kappa\right)-\overline{g}\left(\nu\right),
\end{aligned}
\end{gather}
with
\begin{gather}
\begin{aligned}
\underline{g}\left(\nu\right)
&\doteq
\ln\left(\Gamma\left(\nu+1\right)\right)+\nu\ln\left(2\right)+\overline{c}_{\nu},\\
\overline{g}\left(\nu\right)
&\doteq
\ln\left(\Gamma\left(\nu+1\right)\right)-\frac{1}{2}\ln\left(2\right)-\underline{c}_{\nu}\ln\left(\underline{c}_{\nu}\right).
\end{aligned}
\end{gather}
We further recall that the Rényi divergence is defined as
\begin{align}
d_{\alpha}\left(\mathbf{y},\mathbf{z}\right)
&=
\frac{\nu}{\alpha-1}\ln\left(\frac{\kappa_{y}^{\alpha}\kappa_{z}^{1-\alpha}}{\kappa_{\alpha}}\right)+\frac{\alpha}{\alpha-1}\ln\left(\frac{I_{\nu}\left(\kappa_{\alpha}\right)}{I_{\nu}\left(\kappa_{y}\right)}\right)-\ln\left(\frac{I_{\nu}\left(\kappa_{\alpha}\right)}{I_{\nu}\left(\kappa_{z}\right)}\right),\\
&=\label{fDIV:EQ-REMINDRENYI}
\frac{\nu}{\alpha-1}\ln\left(\frac{\kappa_{y}^{\alpha}\kappa_{z}^{1-\alpha}}{\kappa_{\alpha}}\right)+\frac{\ln\left(I_{\nu}\left(\kappa_{\alpha}\right)\right)}{\alpha-1}-\frac{\alpha\ln\left(I_{\nu}\left(\kappa_{y}\right)\right)}{\alpha-1}-\frac{\left(1-\alpha\right)\ln\left(I_{\nu}\left(\kappa_{z}\right)\right)}{\alpha-1},
\end{align}
and we focus on the final three terms of \cref{fDIV:EQ-REMINDRENYI}.
To bound the final three terms of \cref{fDIV:EQ-REMINDRENYI}, we replace each logarithm with one of the two bounds in \cref{fDIV:EQ-BOUNDDEFINITION}.
Which bound we use depends upon whether we are interested in bounding the Rényi divergence from below or from above, as well as whether the logarithm is added or subtracted, which itself depends upon the magnitude of $\alpha$ (specifically, whether $\alpha$ is less than or greater than one).
Applying this strategy to the final three terms of \cref{fDIV:EQ-REMINDRENYI}, we obtain 
\begin{equation}
\label{fDIV:EQ-RENYILINEARL}
d_{\alpha}\left(\mathbf{y},\mathbf{z}\right)
\geq
\frac{\underline{c}_{\nu}}{\alpha-1}\ln\left(\frac{\kappa_{y}^{\alpha}\kappa_{z}^{1-\alpha}}{\kappa_{\alpha}}\right)+\frac{1}{\alpha-1}\left(g_{\alpha}+\frac{\max\left(\alpha,1\right)}{\mathrm{sign}\left(\alpha-1\right)}\left(\overline{g}\left(\nu\right)-\underline{g}\left(\nu\right)\right)\right),
\end{equation}
and
\begin{equation}
\label{fDIV:EQ-RENYILINEARU}
d_{\alpha}\left(\mathbf{y},\mathbf{z}\right)
\leq
\frac{\underline{c}_{\nu}}{\alpha-1}\ln\left(\frac{\kappa_{y}^{\alpha}\kappa_{z}^{1-\alpha}}{\kappa_{\alpha}}\right)+\frac{1}{\alpha-1}\left(g_{\alpha}+\frac{\max\left(\alpha,1\right)}{\mathrm{sign}\left(\alpha-1\right)}\left(\underline{g}\left(\nu\right)-\overline{g}\left(\nu\right)\right)\right).
\end{equation}
In view of \cref{fDIV:EQ-RENYIDIFFIMPLICATION}, \cref{fDIV:EQ-RENYILINEARL,fDIV:EQ-RENYILINEARU} are identical up to additive constants.
As such, the limiting behaviour of the Rényi divergence is wholly determined by the first term on the right-hand side of \cref{fDIV:EQ-RENYILINEARL,fDIV:EQ-RENYILINEARU}, which is common to both.
\cref{fDIV:EQ-RENYIDIFFEFFECT} implies that the aforementioned first term is a non-trivial function of $\kappa_{y}$. 
It is then trivial to show that the Rényi divergence is an $\mathrm{O}\left(\ln\left(\kappa_{y}\right)\right)$ function.
\end{proof}
\begin{proof}[$\hookrightarrow$ Proof of \cref{fDIV:COR-CSD}]
We focus on the integrand in the definition of the $\chi$-square distance, and note that
\begin{equation}
\label{fDIV:EQ-CSDINTEGRAND}
\frac{f_{p}\left(\mathbf{x};\kappa_{y},\boldsymbol{\mu}_{y}\right)^{2}}{f_{p}\left(\mathbf{x};\kappa_{z},\boldsymbol{\mu}_{z}\right)}
=
\frac{\exp\left(\left(2\kappa_{y}\boldsymbol{\mu}_{y}-\kappa_{z}\boldsymbol{\mu}_{z}\right)'\mathbf{x}\right)}{C_{\nu}\left(\kappa_{y}\right)^{2}/C_{\nu}\left(\kappa_{z}\right)}.
\end{equation}
We then define 
\begin{gather}
\label{fDIV:EQ-KAMUCS}
\begin{aligned}
\kappa_{\chi}
&=
\left\|2\kappa_{y}\boldsymbol{\mu}_{y}-\kappa_{z}\boldsymbol{\mu}_{z}\right\|_{2},\\
\boldsymbol{\mu}_{\chi}
&=
\left\|2\kappa_{y}\boldsymbol{\mu}_{y}-\kappa_{z}\boldsymbol{\mu}_{z}\right\|_{2}^{-1}\left(2\kappa_{y}\boldsymbol{\mu}_{y}-\kappa_{z}\boldsymbol{\mu}_{z}\right).
\end{aligned}
\end{gather}
We substitute these definitions into \cref{fDIV:EQ-CSDINTEGRAND} and integrate, finding that
\begin{equation}
\label{fDIV:EQ-CSDSIMPLE}
\int_{\mathbb{S}^{p-1}}\frac{\exp\left(\kappa_{\chi}\boldsymbol{\mu}_{\chi}'\mathbf{x}\right)}{C_{\nu}\left(\kappa_{y}\right)^{2}/C_{\nu}\left(\kappa_{z}\right)}\,\mathrm{d}\mathbf{x}-1
=
\frac{C_{\nu}\left(\kappa_{\chi}\right)C_{\nu}\left(\kappa_{z}\right)}{C_{\nu}\left(\kappa_{y}\right)C_{\nu}\left(\kappa_{y}\right)}-1.
\end{equation}
Hence, \cref{fDIV:EQ-CSDINTEGRAND,fDIV:EQ-CSDSIMPLE} imply that 
\begin{align}
d_{\chi}\left(\mathbf{y},\mathbf{z}\right)
=
\frac{\kappa_{y}^{p-2}I_{\nu}\left(\kappa_{\chi}\right)I_{\nu}\left(\kappa_{z}\right)}{\kappa_{\chi}^{\nu}\kappa_{z}^{\nu}I_{\nu}\left(\kappa_{y}\right)^{2}}-1,
\end{align}
which is \cref{fDIV:EQ-CHISQ1}. 
The final expression of the $\chi$-square distance can also be obtained via \cref{fDIV:EQ-RENYITOCSD}, with this relationship easily verified.

In the special case where $\kappa_{\chi}=0$, \cref{fDIV:EQ-CSDSIMPLE} reduces to
\begin{equation}
\int_{\mathbb{S}^{p-1}}\frac{1}{C_{\nu}\left(\kappa_{y}\right)^{2}/C_{\nu}\left(\kappa_{z}\right)}\,\mathrm{d}\mathbf{x}-1
=
\frac{2\pi^{\nu+1}/\Gamma\left(\nu+1\right)}{C_{\nu}\left(\kappa_{y}\right)^{2}/C_{\nu}\left(2\kappa_{y}\right)}-1,\\
\end{equation}
which in turn means that 
\begin{equation}
\label{fDIV:EQ-CHISQKEY}
d_{\chi}\left(\mathbf{y},\mathbf{z}\right)
=
\frac{\kappa_{y}^{\nu}I_{\nu}\left(2\kappa_{y}\right)/\Gamma\left(\nu+1\right)}{2^{2\nu}I_{\nu}\left(\kappa_{y}\right)^{2}}-1.
\end{equation}
Here, we use the fact that $\kappa_{\chi}=0$ implies that $2\kappa_{y}=\kappa_{z}$.
In particular, we rely on the fact that
\begin{equation}
C_{\nu}\left(2\kappa_{y}\right)
=
\frac{2\pi^{\nu+1}I_{\nu}\left(2\kappa_{y}\right)}{\kappa_{y}^{\nu}},
\end{equation}
which facilitates much of the cancellation.
We then obtain \cref{fDIV:EQ-CHI} upon rearrangement of \cref{fDIV:EQ-CHISQKEY}.

In the special case where the reference distribution is the uniform distribution on the hypersphere, the integrand in the definition of the $\chi$-square distance has the specific form 
\begin{equation}
\label{fDIV:EQ-CHISQUNIFORM}
\lim_{\kappa_{z}\rightarrow 0}\frac{f_{p}\left(\mathbf{x};\kappa_{y},\boldsymbol{\mu}_{y}\right)^{2}}{f_{p}\left(\mathbf{x};\kappa_{z},\boldsymbol{\mu}_{z}\right)}
=
\frac{2\pi^{\nu+1}\exp\left(2\kappa_{y}\boldsymbol{\mu}_{y}'\mathbf{x}\right)/\Gamma\left(\nu+1\right)}{C_{\nu}\left(\kappa_{y}\right)^{2}}.
\end{equation}
Integrating, we obtain
\begin{equation}
\label{fDIV:EQ-CHISQUNIFORMINTEGRAL}
\int_{\mathbb{S}^{p-1}}\frac{2\pi^{\nu+1}\exp\left(2\kappa_{y}\boldsymbol{\mu}_{y}'\mathbf{x}\right)/\Gamma\left(\nu+1\right)}{C_{\nu}\left(\kappa_{y}\right)^{2}}\,\mathrm{d}\mathbf{x}-1
=
\frac{2\pi^{\nu+1}/\Gamma\left(\nu+1\right)}{C_{\nu}\left(\kappa_{y}\right)^{2}/C_{\nu}\left(2\kappa_{y}\right)}-1,
\end{equation}
but this coincides with \cref{fDIV:EQ-CHISQKEY} and so the expressions of the $\chi$-square distance in the two special cases must coincide.

We now turn our attention to the limiting behaviour of the $\chi$-square distance with respect to an increase in $\kappa_{y}$ given that $\kappa_{y}\geq\kappa_{z}$, holding $\kappa_{z}>0$ fixed.
Although the proof of \cref{fDIV:COR-RD} can be used as a prototype for the construction of a proof here, we instead choose to exploit \cref{fDIV:EQ-RENYITOCSD}.
Since \cref{fDIV:COR-RD} establishes that the Rényi divergence is an $\mathrm{O}\left(\ln\left(\kappa_{y}\right)\right)$ function, the $\chi$-square distance is necessarily an $\mathrm{O}\left(\kappa_{y}\right)$ function. 
\end{proof}
\begin{proof}[$\hookrightarrow$ Proof of \cref{fDIV:COR-HD}]
We focus on the integrand in the definition of the squared-Hellinger distance, and note that
\begin{equation}
\label{fDIV:EQ-HELINTEGRAND}
\sqrt{f_{p}\left(\mathbf{x};\kappa_{y},\boldsymbol{\mu}_{y}\right)f_{p}\left(\mathbf{x};\kappa_{z},\boldsymbol{\mu}_{z}\right)}
=
\frac{\exp\left(\left(\kappa_{y}\boldsymbol{\mu}_{y}+\kappa_{z}\boldsymbol{\mu}_{z}\right)'\mathbf{x}/2\right)}{\sqrt{C_{\nu}\left(\kappa_{y}\right)C_{\nu}\left(\kappa_{z}\right)}}.
\end{equation}
We then define 
\begin{gather}
\label{fDIV:EQ-KAMUHEL}
\begin{aligned}
\kappa_{h}
&=
\frac{1}{2}\left\|\kappa_{y}\boldsymbol{\mu}_{y}+\kappa_{z}\boldsymbol{\mu}_{z}\right\|_{2},\\
\boldsymbol{\mu}_{h}
&=
\left\|\kappa_{y}\boldsymbol{\mu}_{y}+\kappa_{z}\boldsymbol{\mu}_{z}\right\|_{2}^{-1}\left(\kappa_{y}\boldsymbol{\mu}_{y}+\kappa_{z}\boldsymbol{\mu}_{z}\right).
\end{aligned}
\end{gather}
We substitute these definitions into \cref{fDIV:EQ-HELINTEGRAND} and integrate, finding that
\begin{equation}
\label{fDIV:EQ-HELSIMPLE}
2\left(1-\int_{\mathbb{S}^{p-1}}\frac{\exp\left(\left(\kappa_{y}\boldsymbol{\mu}_{y}+\kappa_{z}\boldsymbol{\mu}_{z}\right)'\mathbf{x}/2\right)}{\sqrt{C_{\nu}\left(\kappa_{y}\right)C_{\nu}\left(\kappa_{z}\right)}}\,\mathrm{d}\mathbf{x}\right)
=
2\left(1-\sqrt{\frac{C_{\nu}\left(\kappa_{h}\right)^{2}}{C_{\nu}\left(\kappa_{y}\right)C_{\nu}\left(\kappa_{z}\right)}}\right).
\end{equation}
Hence, \cref{fDIV:EQ-HELINTEGRAND,fDIV:EQ-HELSIMPLE} imply that 
\begin{equation}
d_{h}\left(\mathbf{y},\mathbf{z}\right)^{2}
=
2\left(1-\sqrt{\frac{\kappa_{y}^{\nu}\kappa_{z}^{\nu}I_{\nu}\left(\kappa_{h}\right)^{2}}{\kappa_{h}^{2\nu}I_{\nu}\left(\kappa_{y}\right)I_{\nu}\left(\kappa_{z}\right)}}\right),
\end{equation}
which is \cref{fDIV:EQ-HEL1}.
The final expression of the squared-Hellinger distance can also be obtained via \cref{fDIV:EQ-RENYITOHEL}, with this relationship easily verified.

In the special case where $\kappa_{h}=0$, \cref{fDIV:EQ-HELSIMPLE} reduces to
\begin{equation}
2\left(1-\int_{\mathbb{S}^{p-1}}\frac{1}{\sqrt{C_{\nu}\left(\kappa_{y}\right)C_{\nu}\left(\kappa_{z}\right)}}\,\mathrm{d}\mathbf{x}\right)
=
2\left(1-\frac{2\pi^{\nu+1}/\Gamma\left(\nu+1\right)}{C_{\nu}\left(\kappa_{y}\right)}\right),\label{fDIV:EQ-HELKEY}
\end{equation}
which in turn means that 
\begin{equation}
\label{fDIV:EQ-HELDKEY}
d_{h}\left(\mathbf{y},\mathbf{z}\right)
=
2\left(1-\frac{\kappa_{y}^{\nu}/2^{\nu}}{I_{\nu}\left(\kappa_{y}\right)\Gamma\left(\nu+1\right)},\right).
\end{equation}
Here, we use the fact that $\kappa_{h}=0$ implies that $\kappa_{y}=\kappa_{z}$.
We then obtain \cref{fDIV:EQ-HEL2} upon rearrangement of \cref{fDIV:EQ-HELDKEY}.

In the special case where the reference distribution is the uniform distribution on the hypersphere, the integrand in the definition of the squared-Hellinger distance has the specific form 
\begin{equation}
\label{fDIV:EQ-HELUNIFORM}
\lim_{\kappa_{z}\rightarrow 0}\sqrt{f_{p}\left(\mathbf{x};\kappa_{y},\boldsymbol{\mu}_{y}\right)f_{p}\left(\mathbf{x};\kappa_{z},\boldsymbol{\mu}_{z}\right)}
=
\sqrt{\frac{\Gamma\left(\nu+1\right)}{2\pi^{\nu+1}C_{\nu}\left(\kappa_{y}\right)}}\exp\left(\kappa_{y}\boldsymbol{\mu}_{y}'\mathbf{x}/2\right).
\end{equation}
As such, \cref{fDIV:EQ-HELSIMPLE} reduces to
\begin{equation}
\label{fDIV:EQ-HELUNIFORMINTEGRAL}
2\left(1-\int_{\mathbb{S}^{p-1}}\sqrt{\frac{\Gamma\left(\nu+1\right)}{2\pi^{\nu+1}C_{\nu}\left(\kappa_{y}\right)}}\exp\left(\kappa_{y}\boldsymbol{\mu}_{y}'\mathbf{x}/2\right)\,\mathrm{d}\mathbf{x}\right)
=
2\left(1-\sqrt{\frac{C_{\nu}\left(\kappa_{y}/2\right)^{2}\Gamma\left(\nu+1\right)}{2\pi^{\nu+1}C_{\nu}\left(\kappa_{y}\right)}}\right).
\end{equation}
We then note that
\begin{equation}
\frac{C_{\nu}\left(\kappa_{y}/2\right)^{2}}{C_{\nu}\left(\kappa_{y}\right)}
=
2^{2\nu}\frac{I_{\nu}\left(\kappa_{y}/2\right)^{2}}{I_{\nu}\left(\kappa_{y}\right)}C_{\nu}\left(\kappa_{y}\right)
=
2^{3\nu}\frac{2\pi^{\nu+1}I_{\nu}\left(\kappa_{y}/2\right)^{2}}{\kappa_{y}^{\nu}I_{\nu}\left(\kappa_{y}\right)},
\end{equation}
which we substitute into \cref{fDIV:EQ-HELUNIFORMINTEGRAL} to obtain \cref{fDIV:EQ-HEL3}.

We now turn our attention to the limiting behaviour of the squared-Hellinger distance with respect to an increase in $\kappa_{y}$ given that $\kappa_{y}\geq\kappa_{z}$, holding $\kappa_{z}>0$ fixed.
Although the proof of \cref{fDIV:COR-RD} can be used as a prototype for the construction of a proof here, we instead choose to exploit \cref{fDIV:EQ-RENYITOHEL}.
Since \cref{fDIV:COR-RD} establishes that the Rényi divergence is an $\mathrm{O}\left(\ln\left(\kappa_{y}\right)\right)$ function, the squared-Hellinger distance is necessarily an $\mathrm{O}\left(1-1/\kappa_{y}\right)$ function. 
\end{proof}
\begin{proof}[$\hookrightarrow$ Proof of \cref{fDIV:COR-KL}]
We focus on the integrand in the definition of the Kullback-Leibler divergence and note that
\begin{equation}
\label{fDIV:EQ-KLINTEGRAND}
\ln\left(\frac{f_{p}\left(\mathbf{x};\kappa_{y},\boldsymbol{\mu}_{y}\right)}{f_{p}\left(\mathbf{x};\kappa_{z},\boldsymbol{\mu}_{z}\right)}\right)f_{p}\left(\mathbf{x};\kappa_{y},\boldsymbol{\mu}_{y}\right)
=
\left(\left(\kappa_{y}\boldsymbol{\mu}_{y}-\kappa_{z}\boldsymbol{\mu}_{z}\right)'\mathbf{x}+\ln\left(\frac{C_{\nu}\left(\kappa_{z}\right)}{C_{\nu}\left(\kappa_{y}\right)}\right)\right)f_{p}\left(\mathbf{x};\kappa_{y},\boldsymbol{\mu}_{y}\right),
\end{equation}
We integrate each term on the right-hand side of \cref{fDIV:EQ-KLINTEGRAND}, finding that 
\begin{equation}
\label{fDIV:EQ-KLINTEGRAL1}
\int_{\mathbb{S}^{p-1}}\left(\kappa_{y}\boldsymbol{\mu}_{y}-\kappa_{z}\boldsymbol{\mu}_{z}\right)'\mathbf{x}f_{p}\left(\mathbf{x};\kappa_{y},\boldsymbol{\mu}_{y}\right)\,\mathrm{d}\mathbf{x}
=
\left(\kappa_{y}\boldsymbol{\mu}_{y}-\kappa_{z}\boldsymbol{\mu}_{z}\right)'\mathbb{E}\left(\mathbf{y}\right),
\end{equation}
and
\begin{equation}
\label{fDIV:EQ-KLINTEGRAL2}
\int_{\mathbb{S}^{p-1}}\ln\left(\frac{C_{\nu}\left(\kappa_{z}\right)}{C_{\nu}\left(\kappa_{y}\right)}\right)f_{p}\left(\mathbf{x};\kappa_{y},\boldsymbol{\mu}_{y}\right)\,\mathrm{d}\mathbf{x}
=
\ln\left(\frac{\kappa_{y}^{\nu}I_{\nu}\left(\kappa_{z}\right)}{\kappa_{z}^{\nu}I_{\nu}\left(\kappa_{y}\right)}\right),
\end{equation}
respectively, which relies on the fact that the density function integrates to one.
Adding \cref{fDIV:EQ-KLINTEGRAL1,fDIV:EQ-KLINTEGRAL2} and substituting the results of \cref{fDIV:COR-EV} -- specifically, the result pertaining to the first moment -- we obtain \cref{fDIV:EQ-KL1}.

In the special case where the reference distribution is the uniform distribution on the hypersphere, the integrand in the definition of the Kullback-Leibler divergence has the specific form 
\begin{equation}
\label{fDIV:EQ-KLUNIFORM}
\lim_{\kappa_{z}\rightarrow 0}\ln\left(\frac{f_{p}\left(\mathbf{x};\kappa_{y},\boldsymbol{\mu}_{y}\right)}{f_{p}\left(\mathbf{x};\kappa_{z},\boldsymbol{\mu}_{z}\right)}\right)f_{p}\left(\mathbf{x};\kappa_{y},\boldsymbol{\mu}_{y}\right)
=
\left(\kappa_{y}\boldsymbol{\mu}_{y}'\mathbf{x}+\ln\left(\frac{2\pi^{\nu+1}}{C_{\nu}\left(\kappa_{y}\right)\Gamma\left(\nu+1\right)}\right)\right)f_{p}\left(\mathbf{x};\kappa_{y},\boldsymbol{\mu}_{y}\right).
\end{equation}
We integrate each term on the right-hand side of \cref{fDIV:EQ-KLUNIFORM}, finding that 
\begin{equation}
\label{fDIV:EQ-KLUNIFORMINTEGRAL1}
\int_{\mathbb{S}^{p-1}}\kappa_{y}\boldsymbol{\mu}_{y}'\mathbf{x}f_{p}\left(\mathbf{x};\kappa_{y},\boldsymbol{\mu}_{y}\right)\,\mathrm{d}\mathbf{x}
=
\kappa_{y}\boldsymbol{\mu}_{y}'\mathbb{E}\left(\mathbf{y}\right),
\end{equation}
and
\begin{equation}
\label{fDIV:EQ-KLUNIFORMINTEGRAL2}
\int_{\mathbb{S}^{p-1}}\ln\left(\frac{2\pi^{\nu+1}}{C_{\nu}\left(\kappa_{y}\right)\Gamma\left(\nu+1\right)}\right)f_{p}\left(\mathbf{x};\kappa_{y},\boldsymbol{\mu}_{y}\right)\,\mathrm{d}\mathbf{x}
=
\ln\left(\frac{\left(\kappa_{y}/2\right)^{\nu}}{I_{\nu}\left(\kappa_{y}\right)\Gamma\left(\nu+1\right)}\right),
\end{equation}
respectively.
Adding \cref{fDIV:EQ-KLUNIFORMINTEGRAL1,fDIV:EQ-KLUNIFORMINTEGRAL2} and substituting the results of \cref{fDIV:COR-EV} -- specifically, the result pertaining to the first moment -- we obtain \cref{fDIV:EQ-KL2}.
In particular, we rely on the fact that the inner product of a single mean direction is one.

We now turn our attention to the limiting behaviour of the Kullback-Leibler divergence with respect to an increase in $\kappa_{y}$ given that $\kappa_{y}\geq\kappa_{z}$, holding $\kappa_{z}>0$ fixed.
Given \cref{fDIV:EQ-ORDERLIMIT,fDIV:COR-RD}, it is perhaps unsurprising that we find that the Kullback-Leibler divergence is also an $\mathrm{O}\left(\ln\left(\kappa_{y}\right)\right)$ function.
To show this, however, requires a slight modification of our approach as compared to the proof of \cref{fDIV:COR-RD}.
Whilst the main idea remains the same (replace the logarithms of modified Bessel functions with their lower or upper bounds and characterise the limiting behaviour), the Kullback-Leibler divergence is also a function of ratios of modified Bessel functions.
We must take this into account when deriving bounds.
We rely extensively on \cref{fDIV:COR-LF}, and recall that the Kullback-Leibler divergence is defined as
\begin{equation}
\label{fDIV:EQ-REMINDKL}
d_{\ell}\left(\mathbf{y},\mathbf{z}\right)
=
\nu\ln\left(\frac{\kappa_{y}}{\kappa_{z}}\right)-\ln\left(\frac{I_{\nu}\left(\kappa_{y}\right)}{I_{\nu}\left(\kappa_{z}\right)}\right)+r_{\nu}\left(\kappa_{y}\right)\left(\kappa_{y}-\kappa_{z}\boldsymbol{\mu}_{z}'\boldsymbol{\mu}_{y}\right).
\end{equation}
We bound the Kullback-Leibler divergence from below, as
\begin{align}
d_{\ell}\left(\mathbf{y},\mathbf{z}\right)
&\geq
\nu\ln\left(\frac{\kappa_{y}}{\kappa_{z}}\right)-U\left(\kappa_{y},\kappa_{z}\right)+\frac{\kappa_{y}^{2}}{\underline{c}_{\nu}+\sqrt{\A\kappa_{y}^{2}+\overline{c}_{\nu}^{2}}},\\
&=
-\ln\left(I_{\nu}\left(\kappa_{z}\right)\right)+\underline{c}_{\nu}\ln\left(\frac{\underline{c}_{\nu}+\sqrt{\A\kappa_{y}^{2}+\underline{c}_{\nu}^{2}}}{\underline{c}_{\nu}+\sqrt{\A\kappa_{z}^{2}+\underline{c}_{\nu}^{2}}}\right)-\underline{g}_{\ell}\left(\boldsymbol{\kappa};\nu\right),
\end{align}
where
\begin{align}
\underline{g}_{\ell}\left(\boldsymbol{\kappa};\nu\right)
&\doteq
\sqrt{\A\kappa_{y}^{2}+\underline{c}_{\nu}^{2}}-\sqrt{\A\kappa_{z}^{2}+\underline{c}_{\nu}^{2}}-\frac{\kappa_{y}^{2}}{\underline{c}_{\nu}+\sqrt{\A\kappa_{y}^{2}+\overline{c}_{\nu}^{2}}},\label{fDIV:EQ-KLLSTEPI}\\
&\leq
\sqrt{\A\kappa_{y}^{2}+\overline{c}_{\nu}^{2}}-\sqrt{\A\kappa_{z}^{2}+\underline{c}_{\nu}^{2}}-\sqrt{\A\kappa_{y}^{2}+\overline{c}_{\nu}^{2}}+\overline{c}_{\nu}.\label{fDIV:EQ-KLLSTEPII}
\end{align}
To move from \cref{fDIV:EQ-KLLSTEPI} to \cref{fDIV:EQ-KLLSTEPII}, we replace $\underline{c}_{\nu}$ with $\overline{c}_{\nu}$ where it is appropriate to do so and apply the formula for the difference of two squares (the other square being zero) to the final term.
It then follows that 
\begin{equation}
d_{\ell}\left(\mathbf{y},\mathbf{z}\right)
\geq
-\ln\left(I_{\nu}\left(\kappa_{z}\right)\right)+\underline{c}_{\nu}\ln\left(\frac{\underline{c}_{\nu}+\sqrt{\A\kappa_{y}^{2}+\underline{c}_{\nu}^{2}}}{\underline{c}_{\nu}+\sqrt{\A\kappa_{z}^{2}+\underline{c}_{\nu}^{2}}}\right)-\overline{c}_{\nu}+\sqrt{\A\kappa_{z}^{2}+\underline{c}_{\nu}^{2}}.
\label{fDIV:EQ-KLBELOW}
\end{equation}
We bound the Kullback-Leibler divergence from above, as
\begin{align}
d_{\ell}\left(\mathbf{y},\mathbf{z}\right)
&\leq
\nu\ln\left(\frac{\kappa_{y}}{\kappa_{z}}\right)-L\left(\kappa_{y},\kappa_{z}\right)+\frac{\kappa_{y}^{2}}{\underline{c}_{\nu}+\left(\kappa_{y}^{2}+\underline{c}_{\nu}^{2}\right)^{\frac{1}{2}}},\\
&=
-\ln\left(I_{\nu}\left(\kappa_{z}\right)\right)+\underline{c}_{\nu}\ln\left(\frac{\underline{c}_{\nu}+\left(\kappa_{y}^{2}+\overline{c}_{\nu}^{2}\right)^{\frac{1}{2}}}{\underline{c}_{\nu}+\left(\kappa_{z}^{2}+\overline{c}_{\nu}^{2}\right)^{\frac{1}{2}}}\right)-\overline{g}_{\ell}\left(\boldsymbol{\kappa};\nu\right),
\end{align}
where
\begin{align}
\overline{g}_{\ell}\left(\boldsymbol{\kappa};\nu\right)
&\doteq
\sqrt{\A\kappa_{y}^{2}+\overline{c}_{\nu}^{2}}-\sqrt{\A\kappa_{z}^{2}+\overline{c}_{\nu}^{2}}-\frac{\kappa_{y}^{2}}{\underline{c}_{\nu}+\sqrt{\A\kappa_{y}^{2}+\underline{c}_{\nu}^{2}}},\label{fDIV:EQ-KLUSTEPI}\\
&\geq
\sqrt{\A\kappa_{y}^{2}+\underline{c}_{\nu}^{2}}-\sqrt{\A\kappa_{z}^{2}+\overline{c}_{\nu}^{2}}-\sqrt{\kappa_{y}^{2}+\underline{c}_{\nu}^{2}}+\underline{c}_{\nu}.\label{fDIV:EQ-KLUSTEPII}
\end{align}
To move from \cref{fDIV:EQ-KLUSTEPI} to \cref{fDIV:EQ-KLUSTEPII}, we replace $\overline{c}_{\nu}$ with $\underline{c}_{\nu}$ where it is appropriate to do so and apply the formula for the difference of two squares (the other square being zero) to the final term.
It then follows that 
\begin{equation}
d_{\ell}\left(\mathbf{y},\mathbf{z}\right)
\leq
-\ln\left(I_{\nu}\left(\kappa_{z}\right)\right)+\underline{c}_{\nu}\ln\left(\frac{\underline{c}_{\nu}+\sqrt{\A\kappa_{y}^{2}+\overline{c}_{\nu}^{2}}}{\underline{c}_{\nu}+\sqrt{\A\kappa_{z}^{2}+\overline{c}_{\nu}^{2}}}\right)-\underline{c}_{\nu}+\sqrt{\A\kappa_{z}^{2}+\overline{c}_{\nu}^{2}}.
\label{fDIV:EQ-KLABOVE}
\end{equation}
Together, \cref{fDIV:EQ-KLBELOW,fDIV:EQ-KLABOVE} imply that
\begin{equation}
\left\vert d_{\ell}\left(\kappa_{y},\kappa_{z}\right)\right\vert
\leq
\underline{c}_{\nu}\ln\left(\frac{\underline{c}_{\nu}+\sqrt{\A\kappa_{y}^{2}+\underline{c}_{\nu}^{2}}}{\underline{c}_{\nu}+\sqrt{\A\kappa_{z}^{2}+\underline{c}_{\nu}^{2}}}\right)+\left\vert\sqrt{\A\kappa_{z}^{2}+\overline{c}_{\nu}^{2}}-\underline{c}_{\nu}-\ln\left(I_{\nu}\left(\kappa_{z}\right)\right)\right\vert,
\end{equation}
and so it is trivial to show that the Kullback-Leibler divergence is an $\mathrm{O}\left(\ln\left(\kappa_{y}\right)\right)$ function.
\end{proof}
\begin{proof}[$\hookrightarrow$ Proof of \cref{fDIV:COR-LF}]
We rewrite \cref{fDIV:EQ-LUNIFORM,fDIV:EQ-UUNIFORM} using the formula for the difference of two squares as 
\begin{align}
L_{\nu}\left(\kappa,0\right)
&=
\frac{1}{2}\ln\left(\frac{2}{\kappa}\right)-\ln\left(\Gamma\left(\nu+1\right)\right)+\underline{c}_{\nu}\ln\left(\frac{\kappa\left(\underline{c}_{\nu}+\overline{c}_{\nu}\right)/2}{\underline{c}_{\nu}+\sqrt{\A\kappa^{2}+\overline{c}_{\nu}^{2}}}\right)+\sqrt{\A\kappa^{2}+\overline{c}_{\nu}^{2}}-\overline{c}_{\nu},\\
&\geq
\frac{1}{2}\ln\left(\frac{2}{\kappa}\right)-\ln\left(\Gamma\left(\nu+1\right)\right)+\underline{c}_{\nu}\ln\left(\frac{\kappa\left(\underline{c}_{\nu}+\overline{c}_{\nu}\right)/2}{\underline{c}_{\nu}+\kappa+\overline{c}_{\nu}}\right)+\kappa-\overline{c}_{\nu},\\
&=
\frac{1}{2}\ln\left(\frac{2}{\kappa}\right)-\ln\left(\Gamma\left(\nu+1\right)\right)+\underline{c}_{\nu}\ln\left(\kappa\right)+\underline{c}_{\nu}\ln\left(\frac{\underline{c}_{\nu}+\overline{c}_{\nu}}{2\left(\underline{c}_{\nu}+\overline{c}_{\nu}\right)+\kappa}\right)+\kappa-\overline{c}_{\nu},\\
&\geq
\frac{1}{2}\ln\left(\frac{2}{\kappa}\right)-\ln\left(\Gamma\left(\nu+1\right)\right)+\left(\underline{c}_{\nu}-\underline{c}_{\nu}\right)\ln\left(\kappa\right)+\underline{c}_{\nu}\ln\left(\frac{\underline{c}_{\nu}+\overline{c}_{\nu}}{2\left(\underline{c}_{\nu}+\overline{c}_{\nu}\right)}\right)+\kappa-\overline{c}_{\nu},\\
&=
\kappa-\frac{1}{2}\ln\left(\kappa\right)-\nu\ln\left(2\right)-\ln\left(\Gamma\left(\nu+1\right)\right)-\overline{c}_{\nu},\label{fDIV:EQ-ASYMPTOTICL}
\end{align}
and
\begin{align}
U_{\nu}\left(\kappa_y,0\right)
&=
\frac{1}{2}\ln\left(\frac{2}{\kappa_y}\right)-\ln\left(\Gamma\left(\nu+1\right)\right)+\underline{c}_{\nu}\ln\left(\frac{\kappa\left(\underline{c}_{\nu}+\underline{c}_{\nu}\right)/2}{\underline{c}_{\nu}+\sqrt{\A\kappa^{2}+\underline{c}_{\nu}^{2}}}\right)+\sqrt{\A\kappa^{2}+\underline{c}_{\nu}^{2}}-\underline{c}_{\nu},\\
&\leq
\frac{1}{2}\ln\left(\frac{2}{\kappa_y}\right)-\ln\left(\Gamma\left(\nu+1\right)\right)+\underline{c}_{\nu}\ln\left(\frac{\underline{c}_{\nu}\kappa}{\kappa}\right)+\kappa+\underline{c}_{\nu}-\underline{c}_{\nu},\\
&=
\kappa-\frac{1}{2}\ln\left(\kappa\right)+\frac{1}{2}\ln\left(2\right)-\ln\left(\Gamma\left(\nu+1\right)\right)+\underline{c}_{\nu}\ln\left(\underline{c}_{\nu}\right),\label{fDIV:EQ-ASYMPTOTICU}
\end{align}
respectively.
Here, we exploit the concavity of the logarithmic and square root functions, and set some terms equal to zero where it is useful to do so.
This establishes the veracity of the first part of \cref{fDIV:COR-LF}.
We now need to show that the logarithm of the modified Bessel function, which we know lies between these two bounds for all $\kappa>0$, is an $\mathrm{O}\left(\kappa-\ln\left(\kappa\right)/2\right)$ function.
To do so, we show that the maximum of the absolute value of \cref{fDIV:EQ-ASYMPTOTICL,fDIV:EQ-ASYMPTOTICU} is bounded from above by a function that has the required order.
Specifically, for all $\kappa>0$,
\begin{align}
|\ln\left(I_{\nu}\left(\kappa\right)\right)|
&\leq
\kappa-\frac{1}{2}\ln\left(\kappa\right)+\max\left(\frac{1}{2},\nu\right)\ln\left(2\right)+\ln\left(\Gamma\left(\nu+1\right)\right)+\max\left(\overline{c}_{\nu},\underline{c}_{\nu}\ln\left(\underline{c}_{\nu}\right)\right),\\
&\leq\label{fDIV:EQ-WHYBLOWUP}
\left(\kappa-\frac{1}{2}\ln\left(\kappa\right)\right)\left(1+\frac{\max\left(1/2,\nu\right)\ln\left(2\right)+\ln\left(\Gamma\left(\nu+1\right)\right)+\max\left(\overline{c}_{\nu},\underline{c}_{\nu}\ln\left(\underline{c}_{\nu}\right)\right)}{\left(1+\ln\left(2\right)\right)/2}\right),
\end{align}
where the denominator in the second term of \cref{fDIV:EQ-WHYBLOWUP} is the minimum value that the first term of \cref{fDIV:EQ-WHYBLOWUP} can attain, which establishes the result.
\end{proof}
\section{Hankel expansion of the circular variance}
\label{fDIV:HANKEL}
One quantity that is often of interest is the circular variance of the von Mises-Fisher family of distributions, which is defined as one minus the ratio of modified Bessel functions -- i.e., one minus the mean resultant length.
\citet{KitagawaLopez} demonstrates that the circular variance is an $\mathrm{O}\left(1/\kappa\right)$ function.
Here, we show that the same result can be obtained via Hankel series expansion (see \cref{fDIV:EQ-HANKEL} for a definition).
We note that Hankel series expansion is appropriate when $\kappa\rightarrow\infty$, which is the limiting behaviour that we are interested in.
We begin by writing the circular variance in the alternative form
\begin{equation}
1-r_{\nu}\left(\kappa\right)
=
1-\frac{I_{\nu+1}\left(\kappa\right)}{I_{\nu}\left(\kappa\right)}
=
\frac{I_{\nu}\left(\kappa\right)-I_{\nu+1}\left(\kappa\right)}{I_{\nu}\left(\kappa\right)},
\end{equation}
to which we apply the expansion.
Recalling \cref{fDIV:EQ-HANKEL,fDIV:EQ-HANKELSERIES},
\begin{align}
1-r_\nu\left(\kappa\right)
&=
\left(1-\frac{a_{1}\left(\nu\right)}{\kappa}+\mathrm{O}\left(\frac{1}{\kappa^{2}}\right)\right)^{-1}\left(\frac{a_{1}\left(\nu+1\right)}{\kappa}-\frac{a_{1}\left(\nu\right)}{\kappa}+\mathrm{O}\left(\frac{1}{\kappa^{2}}\right)\right),\\
&=
\frac{a_{1}\left(\nu+1\right)-a_{1}\left(\nu\right)+\mathrm{O}\left(\kappa\right)}{\kappa-a_{1}\left(\nu\right)+\mathrm{O}\left(\kappa\right)},
\end{align}
which is, trivially, an $\mathrm{O}\left(1/\kappa\right)$ function.
\renewcommand*\thesection{\arabic{section}}
\bibliography{BIBFILE.bib}
\end{document}